\documentclass[11pt]{article}
\usepackage{bailey}
\usepackage{booktabs} 
\usepackage{fullpage}
\usepackage[mathscr]{euscript}
\usepackage{comment}
\usepackage{enumitem}
\usepackage[table]{xcolor}
\usepackage[colorlinks = true,allcolors = teal]{hyperref}
\usepackage{cleveref}
\usepackage{nicefrac}
\usepackage{url}
\usepackage{multirow}
\usepackage{wrapfig}
\usepackage[numbers]{natbib}

\newcommand{\emdash}{\,---\,}
\newcommand{\profiles}{\Pi}
\newcommand{\profile}{{\boldsymbol{\pi}}}
\newcommand{\cands}{M}
\newcommand{\rankings}{\mathcal{L}}
\newcommand{\ranking}{\pi}
\newcommand{\notshortrankings}{\rankings(\cands)}
\newcommand{\noisemodel}{\mathcal{S}}
\newcommand{\proportions}{\mathbf{h}}
\newcommand{\proportion}{h}

\newcommand{\E}{\mathbb{E}}

\newcommand{\prop}{\mathscr{H}}
\newcommand{\minprob}{\textsc{min-prob}}
\newcommand{\suchthat}{\ : \ }
\newcommand{\conditioned}{\ | \ }
\newcommand{\set}[1]{{\left\{#1\right\}}}
\newcommand{\event}{\mathcal{E}}
\newcommand{\rankingsmissing}{\rankings_{-1}}
\newcommand{\rev}{\text{rev}}

\DeclareMathOperator*{\argmax}{argmax}

\newcommand{\rankingdistranking}{\noisemodel_\phi(\ranking)}
\newcommand{\rankingdistrankingi}{\noisemodel_\phi(\ranking_i)}
\newcommand{\rankingdistprof}{\noisemodel_\phi(\profile)}

\newcommand{\propdistranking}{\prop(\rankingdistranking)}
\newcommand{\propdistrankingi}{\prop(\rankingdistrankingi)}
\newcommand{\propdistprof}{\prop(\rankingdistprof)}

\newcommand{\hq}{H^{\mathbb{Q}}}

\allowdisplaybreaks

\newtheoremstyle{sctheorem}%
{}{}%
{\addtolength{\leftskip}{0em}}
{}%
{\scshape}{.}%
{ }%
{\thmname{#1}\thmnumber{ #2}\thmnote{{\normalfont\ (#3)}}}
\theoremstyle{sctheorem}

\newtheorem{theorem}{Theorem}

\newtheorem{corollary}[theorem]{Corollary}

\theoremstyle{definition}

\newtheorem{example}[theorem]{Example}

\theoremstyle{sctheorem}
\newtheorem{assumption}{Assumption}
\newtheorem{lemma}[theorem]{Lemma}
\newtheorem{definition}[theorem]{Definition}

\newtheorem{proposition}[theorem]{Proposition}

\crefname{assumption}{Assumption}{Assumptions}

\newcommand{\e}{\varepsilon}

\title{Smoothed Analysis of Social Choice, Revisited}

\author{Bailey Flanigan \\ Carnegie Mellon University \\ \texttt{\small{bflaniga@andrew.cmu.edu}}
 \and Daniel Halpern \\ Harvard University \\ \texttt{\small{dhalpern@g.harvard.edu}} \and Alexandros Psomas \\ Purdue University \\ \texttt{\small{apsomas@cs.purdue.edu}}}

\date{}

\begin{document}

\maketitle

\begin{abstract}
A canonical problem in social choice is how to aggregate ranked votes: that is, given $n$ voters' rankings over $m$ candidates, what \textit{voting rule} $f$ should we use to aggregate these votes and select a single winner? One standard method for comparing voting rules is by their satisfaction of \textit{axioms}\emdash properties that we want a ``reasonable'' rule to satisfy. This approach, unfortunately, leads to several impossibilities: no voting rule can simultaneously satisfy all the properties we would want, at least in the worst case over all possible inputs.

Motivated by this, we consider a relaxation of this worst case requirement. We analyze this through a ``smoothed'' model of social choice, where votes are (independently) perturbed with small amounts of noise. If no matter which input profile we start with, the probability (post-noise) of an axiom being satisfied is large, we will view it as nearly as good as satisfied\emdash called ``smoothed-satisfied''\emdash even if it may be violated in the worst case.

Our model is a mild restriction of Lirong Xia's, and corresponds closely to that in Spielman and Teng's original work on smoothed analysis. Much work has been done so far in several papers by Xia on which axioms can and cannot be satisfied under such noise. In our paper, we aim to give a more cohesive overview on when smoothed analysis of social choice is useful. 

Within our model, we give simple sufficient conditions for smoothed-satisfaction or smoothed-violation of several axioms and paradoxes, including most of those studied by Xia as well as some previously unstudied. We then observe that, in a practically important subclass of noise models, although convergence eventually occurs, known rates may require an extremely large number of voters. Motivated by this, we prove bounds specifically within a canonical noise model from this subclass\emdash the Mallows model. Here, we find a more nuanced picture on exactly when smoothed analysis an help. 
\end{abstract}

\section{Introduction} \label{sec:introduction}
One of the most canonical problems in social choice is how to aggregate votes: that is, given a \textit{preference profile} consisting of $n$ voters' rankings over $m$ alternatives, what \textit{voting rule} $f$ should we use to aggregate this preference profile into a single winner? A predominant way of comparing voting rules is by their satisfaction of \textit{axioms}\emdash logical statements that describe basic, natural properties that voting rules should satisfy. 

Traditionally, it is said that a voting rule \textit{satisfies} an axiom if the logical statement specified by the axiom holds on \textit{all possible} preference profiles; if there exists even one profile on which it does not hold (a \textit{counterexample}),  
the rule is said to \textit{violate} the axiom. Unfortunately, under this worst-case notion of satisfaction, \textit{all} %
voting rules violate at least some common-sense axioms.
\footnote{E.g., the Gibbard-Satterthwaite theorem says that any onto, single-winner, non-dictatorial voting rule with $m > 2$ is not strategyproof~\cite{gibbard1973manipulation}. For more examples, see the \href{https://en.wikipedia.org/wiki/Comparison_of_electoral_systems}{Comparison of electoral systems} Wikipedia page.
}
There is hope, however, because this worst-case approach may cover up an important distinction between voting rules: while some rules may fail to satisfy an axiom on large swaths of preference profiles, others may fail only on precisely contrived instances. This distinction is of practical significance because voting rules of the latter type should usually satisfy the axiom in practice, where the lack of perfect correlation between people's preferences makes extremely contrived counterexamples unlikely to occur. 

In 2020, a paper by Lirong Xia aimed to capture this intuition by modeling profiles as \textit{semi-random}\emdash mostly adversarial, but with random perturbations~\cite{xia2020smoothed}. More precisely, Xia's model assumes that ``noisy'' (but otherwise adversarial) profiles arise in the following way: first, let an adversary choose voters' ``types,'' with each type corresponding to a distribution over rankings. The ultimate profile is then sampled by selecting each voter's ranking independently from their type distribution. 

In this paper, we capture the same intuition using a slightly restricted version of Xia's semi-random model. Our model produces ``noisy'' profiles in the following way: first, let an adversary fix a \textit{starting profile}. Then, apply a small amount of noise independently to each individual ranking in the profile\emdash that is, for each ranking, draw a new ranking from some distribution (this distribution can be arbitrary, up to regularity conditions). We will refer to this as the \textit{smoothed} model, as it is directly analogous to Spielman and Teng's ``smoothed'' model in the distinct context of linear programming. We discuss this connection and the precise relationship between our model and Xia's in \Cref{sec:related}. The key takeaway is that ours is a slight restriction, assumed primarily for ease of exposition.

Because the smoothed model goes so minimally beyond the worst case, resolving axioms in this way is quite meaningful: the small amount of noise it assumes should be sufficient to escape counterexamples only if they are truly isolated among other profiles, i.e., contrived. The model is also well-motivated practically, as people's preferences are established to be susceptible to small shocks, in daily life and even in the preference elicitation process~\cite{dillman2005survey,kahneman2013prospect,lee2009search}. The existing work in this area gives some hope of such positive results: it shows 
that under all standard voting rules considered, the axioms Group-strategyproofness~\cite{xiamanipulation}, Resolvability~\cite{xia2021likely}, and Participation~\cite{xia2021axioms} are satisfied post-noise with high probability as $n$ grows large. In contrast, the axiom Condorcet consistency remains violated with high probability, even after semi-random noise, by a popular class of rules~\cite{xia2021axioms}. Here we see heterogeneity across axioms: for some, smoothed noise reliably circumvents impossibilities, whereas, for others, it seems insufficient. This motivates our first question:\\[-0.5em]

\noindent \textbf{Question 1:} \textit{What properties of axiomatic impossibilities make them resolvable by smoothed noise?}\\[-0.5em]

This question remains to be fully addressed, in part because existing work has mostly analyzed axioms one by one, 
using technically intricate arguments to get precise asymptotic convergence rates in $n$. While this approach has yielded detailed insights about specific axioms and voting rules, it remains to distill higher-level patterns across rules and axioms.

Our second research question is a refinement of the first and deals with understanding when semi-random noise can circumvent impossibilities, \textit{absent a key assumption made in the related work so far}. This assumption is called \textit{positivity} in the related work, and it amounts to assuming that when a ranking is perturbed, there is at least some fixed probability, $\minprob$, of \textit{any other ranking} resulting from this perturbation. Although it is not explicitly written, existing upper bounds on rates of convergence for the axioms we consider implicitly depend multiplicatively on $1/\minprob^{3/2}$. Although when $\minprob$ is large, this polynomial dependence is not too demanding, when $\minprob$ is small, enormous numbers of voters may be necessary to guarantee a reasonable probability of axiom satisfaction. 

Although assuming that $\minprob$ is large enough to avoid such issues may seem innocuous, it is anything but. This is because noise models in which $\minprob$ is very small represent perhaps the most realistic subclass of noise models: they encompass any noise model in which small perturbations are unlikely to \textit{completely reverse} someone's ranking (or make similarly drastic changes). Given that small real-world shocks that may perturb people's preferences are unlikely to qualitatively change their opinions (to, e.g., the extent that their rankings are reversed), a more realistic noise model might be one in which ``less'' noise corresponds to lower probability of drawing a perturbed ranking that reflects extreme opinion change. This motivates our second research question:\\[-0.5em]
 
\noindent \textbf{Question 2:} \textit{What properties of axiomatic impossibilities make them resolvable by smoothed noise, for noise models in which rankings reflecting extreme opinion change are drawn with low (or zero) probability?}

\subsection{Results and Contributions}
\begin{wrapfigure}{r}{0.5\textwidth}
\vspace{-4em}
\centering
    \includegraphics[width=0.5\textwidth]{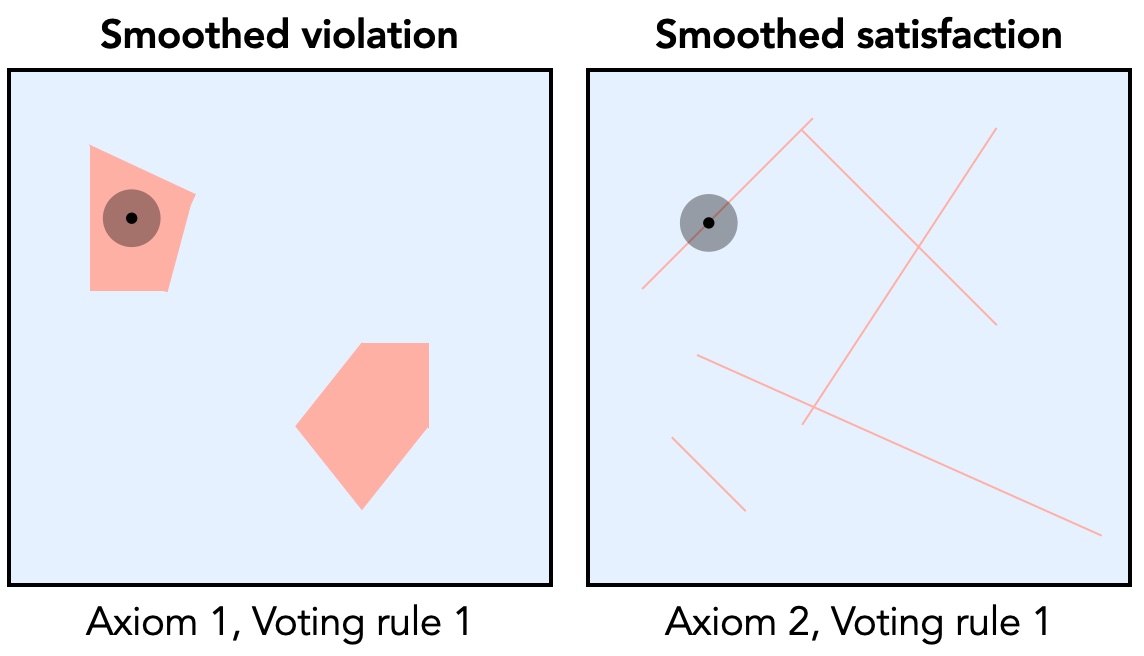}
  \caption{Both panes represent (abstractly) the space of all possible profiles. 
  Red profiles are counterexamples in which the voting rule fails to satisfy the axiom. The black dot is the starting profile, and surrounding region is where the resulting profile is likely to end up after applying noise.}\label{fig:fig1}
    \vspace{-1em}
\end{wrapfigure}
\paragraph{Question 1.}
In \Cref{sec:results1}, we study the general smoothed model. We distill two patterns: one across axioms where smoothed noise \textit{does} help, and another across axioms where it does not. In the process, we do the first smoothed (or semi-random) analysis of several core axioms in social choice. We also re-analyze some axioms studied in past work, but we prove these results anew, with derivations that are standardized across axioms and yield bounds that explicitly depend on $\minprob$, which will be useful in Question 2. We delineate which analyses are new and which are re-proven in \Cref{sec:related}.

As summarized in \Cref{tab:classification} (located in \Cref{sec:negative-results}), we first show negative results for the axioms \textsc{Condorcet Consistency}, \textsc{Independence of Irrelevant Alternatives}, \textsc{Consistency}, \textsc{Majority}. Across all voting rules we study, we find that these axioms are all smoothed-violated whenever they are violated, i.e., smoothed noise is insufficient to circumvent impossibilities for these axioms. Moreover, across axioms and rules, this smoothed violation always holds for the same reason: there exist large contiguous segments of the profile space consisting of counterexamples.
As shown in the left-hand pane of \Cref{fig:fig1}, if such regions exist, then our starting profile can be chosen to be inside one, in which case small perturbations are insufficient to produce profiles outside it. We formalize this underlying pattern into a generic sufficient condition for smoothed-violation in \Cref{thm:violation-sufficient}.

As summarized in \Cref{tab:classification2} (located in \Cref{sec:positive-results}), we next show positive results for the axioms \textsc{Resolvability} and group notions of \textsc{Strategyproofness}, \textsc{Participation}, and \textsc{Monotonicity}. In the spirit of generalizing across axioms, we are able to analyze the latter three at once by proving the smoothed-satisfaction of a more general axiom, called $\rho(n)$-\textsc{Group-Stability}, which requires that the behavior of $\rho(n)$ voters should not be able to affect the outcome of the election (essentially framing \emph{Margin of Victory} as a binary property~\cite{pritchard2006average,xiamanipulation}). The common feature of these axioms, which permits their smoothed-satisfaction, is that their counterexamples must occur on or near hyperplanes. As depicted in the right-hand pane of Figure 1, this means that any ``well-behaved'' noise, applied to any starting profile, will place very little probability mass on the sliver of counterexamples contained within (potentially some nonzero but shrinking distance of) the measure-zero hyperplane. We formalize this underlying pattern into a generic sufficient condition for smoothed-satisfaction in \Cref{thm:brittle_convex}.

\paragraph{Question 2.} 
In \Cref{sec:beyond-minprob}, we pursue the first convergence rates to smoothed satisfaction that are not parameterized by $\minprob$, and which do not rely on the minimum probability being nonzero. Although the core ideas we use naturally extend to broader classes of noise models, for concreteness, we pursue these bounds specifically in the \textit{Mallows model}\emdash perhaps the most canonical model of noisy rankings. This model has the realistic property we want: the probability of drawing a ranking decreases in its \textit{Kendall tau distance}\footnote{The Kendall tau distance between two rankings $\ranking,\ranking'$ is the number of pairs of candidates on which they disagree.} from the original ranking. More precisely, for a noise parameter $\phi \in [0,1]$, the probability of drawing a ranking at Kendall tau distance $d$ from the original ranking is proportional to $\phi^{d}$. As a result, $\minprob$ for this model is exponentially small: the probability of reversing a ranking due to Mallows noise is $\phi^{\Omega(m^2)}$.

The Mallows model falls within our smoothed model, so our impossibilities from \Cref{sec:negative-results} carry over. From among the remaining axioms of interest (those in \Cref{sec:positive-results}), we characterize precise convergence rates to smoothed satisfaction of \textsc{Resolvability} and $\rho(n)$\textsc{-Group-Strategyproofness} in the Mallows model. 

While $\minprob$-parameterized upper bounds yielded heterogeneity across axioms, they were almost homogeneous across voting rules per axiom, i.e., for each axiom, either nearly all rules satisfied, or nearly all rules violated them. Interestingly, these new bounds reveal diversity across voting rules as well. As summarized in \Cref{tab:classification4}, while \textsc{Plurality} and \textsc{Borda Count}
are smoothed-satisfied at a rate polynomial in $\phi$ and $m$, we can lower bound the convergence rates of \textsc{Maximin} and  \textsc{Veto} to include a $1/\phi^{\Omega(m)}$ term. This essentially means the number of voters required to guarantee satisfaction is exponential in these parameters, making this property less appealing.  Across these voting rules, we can distill a pattern: voting rules whose outcomes are sensitive to changes throughout ranking positions are helped substantially by Mallows noise, while those whose outcomes remain fixed despite potentially many swaps are helped much less. 
Zooming out, this conceptual finding extends beyond the Mallows model and the studied axioms, suggesting that for reasonable noise models, only voting rules that are truly sensitive to small changes in voters' preferences will have brittle impossibilities.

\paragraph{Bonus Contribution: smoothed analysis of Arrow's theorem.} As an extension, in \Cref{sec:discussion} we do the first smoothed (or semi-random) analysis of \textit{Arrow's Theorem} \cite{arrow1951individual}\emdash perhaps one of the most influential impossibilities in social choice. Surprisingly, we show that this impossibility is resolved with high probability under the smoothed model. However, as we show, this resolution is for a trivial reason related to Arrow's definition of the axiom \textit{Non-dictatorship}. We therefore identify a slight strengthening of Arrow's theorem, and pose the open question of whether smoothed noise is sufficient to resolve it.

\subsection{Related work}\label{sec:related}
Our model is designed to be a close analog to Spielman and Teng's celebrated ``smoothed'' model, introduced in their analysis of the Simplex algorithm \cite{spielman2004smoothed}. We give a detailed description of the parallels between our model and theirs in \Cref{app:st_model}. Within social choice, the model most closely related to ours is Lirong Xia's semi-random model \cite{xia2020smoothed},\footnote{To avoid confusion with our model, we clarity: Xia's model is called ``smoothed'' in the original paper, but is renamed ``semi-random'' in most subsequent work.} of which our model is a mild restriction. To see how Xia's model generalizes ours, in our model we always have $m!$ ``types'', with each type corresponding to a possible starting ranking. Each type's associated distribution is then the noise added to that ranking. The key restriction we make is that, while Xia's model allows types with potentially different ``shapes'' of noise distributions, we assume that all rankings are perturbed by a common noise distribution, just centered at different rankings. 
This restriction is technically mild, as the assumption critical to analyses in both models is that noise is \textit{independent} across rankings, and many of our results could be expressed in the more general model but in a less concise way. Because of that, we chose to stick to a slightly more restricted model for ease of exposition.

Beyond Xia's model, axiom satisfaction has also been studied in other, less-adversarial randomized models. One popular such model is \textit{Impartial Culture}, where profiles are drawn uniformly and i.i.d.~\cite{gehrlein2002condorcet,green2016statistical,mossel2022probabilistic}. Slightly more generally, the axiom Strategyproofness was studied in a \textit{non-uniform} i.i.d model by \citet{mossel2013smooth}. Further afield, there is empirical work on the frequency of axiom violations on simulated inputs and real elections \cite{plassmann2014frequently,felsenthal1993empirical}.

Now we situate our results among past work on axiom satisfaction in the semi-random model. First, our results in \Cref{sec:beyond-minprob} are completely distinct from this existing line of work, pursuing bounds that do not rely on the positivity assumption core to the semi-random model. In contrast, our results in \Cref{sec:results1} are for generic noise distributions that \textit{are} subject to the positivity assumption, along with all other regularity conditions imposed in Xia's semi-random model. In the general smoothed model, we present the first smoothed (or semi-random) analysis of the axioms \textsc{Consistency}, \textsc{Independence of Irrelevant Alternatives}, and \textsc{Majority}, along with Arrow's impossibility. We also introduce and analyze a new axiom, $\rho(n)$-\textsc{Group-Stability}, which we show generalizes multiple standard axioms. In the interest of identifying patterns, we also re-analyze several axiomatic impossibilities for which there already exist similar bounds,\footnote{These axiomatic results include \textsc{Resolvability} \cite{xia2021likely}; \textsc{Strategyproofness} \cite{xiamanipulation}; \textsc{Condorcet} and \textsc{Participation} \cite{xia2021axioms}, plus Moulin's impossibility of satisfying them together~\cite{xia2021smoothed}; and Condorcet's voting paradox and the ANR impossibility theorem \cite{xia2020smoothed}. } though we show them via different proofs and give bounds that are explicit in their dependence on $\minprob$.
For clarity, we distinguish axioms and impossibilities analyzed for the first time in this paper in \Cref{tab:classification}, \Cref{tab:classification2}, and \Cref{tab:classification4} with $^*$.
The set of voting rules we study overlaps\emdash but not perfectly\emdash with the existing work on these axioms. In \Cref{app:techniques}, we offer a detailed comparison between the techniques we use to prove our bounds versus those used in past work.

\section{Model} \label{sec:model}

\noindent \textbf{Rankings.} There are $m$ candidates $\cands$ and $n$ voters $N$. Voters express their preferences over the candidates as complete rankings. Formally, a ranking $\ranking$ is a bijection from indices to candidates $[m] \to \cands$, where $\ranking(j)$ represents the candidate ranked $j$-th in $\ranking$. Let $a \succ_\ranking b$ to denote that candidate $a$ is preferred to $b$ in ranking $\ranking$, or formally, $\ranking^{-1}(a) < \ranking^{-1}(b)$. 
$\notshortrankings$ is the set of $m!$ possible rankings, or just $\mathcal{L}$ when $M$ is clear from context. 
Letting the rankings in $\rankings$ be implicitly ordered, the last ($m!$-th) ranking in $\rankings$ is $\ranking_{-1}$, and the set of rankings without this element as $\rankingsmissing = \rankings \setminus \{\ranking_{-1}\}$. For reasons to be clarified next, we will often work with $\rankingsmissing$ instead of $\rankings$.
\\[-0.5em]

\noindent\textbf{Profiles and profile histograms}. A profile $\profile = (\ranking_i | i\in N)$ is a vector of $n$ rankings, where $\ranking_i$ is voter $i$'s ranking. 
Let $\profiles_n = \prod_{i = 1}^n \rankings$ be the set of all profiles on $n$ voters, and let $\profiles = \bigcup_{n \in \mathbb{Z}^+} \profiles_n$ be the set of all profiles overall. We define addition over profiles in the natural way: $(\profile + \profile') = (\ranking_i | i\in N) \| (\ranking_i' | i \in N')$.\footnote{Here $\|$ represents concatenation, and thus the resulting profile $\profile + \profile'$ contains $|N| + |N'|$ voters. 
We also extend addition to permit positive integer multiples of profiles: $z \profile$ means adding a profile together $z$ times.} 

Instead of working directly with profiles, we will work primarily with their \textit{histograms}. A \textit{histogram} $\proportions$ is an $|\rankingsmissing|$-length vector whose $\ranking$-th entry $\proportion_{\ranking}$ is the proportion of voters with ranking $\ranking$ in a profile ($\proportions$ is indexed only up to $\rankingsmissing$ because the histogram must add to 1, so the $m!$-th index is redundant). 
The histogram associated with a particular profile $\profile$ is $\proportions^\profile$, with $\pi$-th entry $\proportions^{\profile}_{\ranking} := \nicefrac{1}{n} \, |\{i \suchthat \ranking_i = \ranking\}|.$ 
We define the simplex of all possible histograms as 
    $\Delta_{\proportions} := \left\{\proportions \in [0,1]^{|\rankingsmissing|} \suchthat \sum_{\ranking \in \rankingsmissing} \proportion_\ranking \le 1\right\}.$
Of course, $H$ includes vectors with irrational entries that could never correspond to a well-defined profile; nonetheless, it will be useful to consider the completed space. In order to talk about only the histograms that are realizable from well-defined profiles, we also define $\hq = H \cap \mathbb{Q}^{|\rankingsmissing|}$ to be the subset of $H$ of vectors with rational components.

Finally, we will use \textit{histogram operator} $\prop(\cdot)$ to transform more general, profile-based objects into their histogram-based analogs. For example, $\prop(\profile) = \proportion^{\profile}$, and $\prop(\profiles) = H^{\mathbb{Q}}$. For a single ranking, $\prop(\ranking)$ is a $|\rankingsmissing|$-length basis vector with a 1 at the $\ranking$-th index ($\prop(\ranking_{-1})$ is the 0s vector). We will also apply this operator to \textit{distributions over} profiles and rankings in the natural way, first drawing a profile or ranking from the distribution, and then considering its corresponding histogram.\\[-0.5em]





\noindent\textbf{Voting Rules.} A \textit{voting rule} is a function $R: \profiles \mapsto 2^{M}$ mapping a given profile to a set of winning candidates. Then, $R(\profile)$ is the set of winners chosen by the voting rule on $\profile$. If a voting rule by its standard definition results in a tie, rather than specifying a tie-breaking rule, we assume it returns all such winners (this assumption is for ease of exposition only). Let $\mathcal{R}$ be the set of all voting rules. We will study several specific rules, defined colloquially below and formally in \Cref{app:rules}.

\textit{Positional Scoring Rules} (PSRs) are represented by $m$-length vectors of weakly decreasing scores $s_1 \geq s_2 \geq \dots \geq s_m$, where $s_1=1$ and $s_m=0$. Candidate $a$ receives $s_i$ points for each voter that ranks it $i$-th; the candidate with the most points wins. We consider three specific PSRs, defined by their score vectors: \textsc{Plurality}: $(1,0,\dots,0,0)$, \textsc{Borda}: $(1,1-\frac{1}{m-1},\dots,\frac{1}{m-1},0)$, and \textsc{Veto}: $(1,1,\dots,1,0)$. Beyond PSRs, we study the rules \textsc{Minimax}, \textsc{Kemeny-Young}, and \textsc{Copeland}. \textsc{Minimax} selects the candidate with the smallest \textit{maximum pairwise domination}.\footnote{Useful definitions: $a$ \textit{pairwise-dominates} $b$ in $\profile$ when over half of voters rank $a$ ahead of $b$: $\big|\{\ranking_i| a \succ_{\ranking_i} b\}\big| > n/2$. $a$ and $b$\textit{ pairwise tie} when $\big|\{\ranking_i| a \succ_{\ranking_i} b\}\big| = n/2$. $a$'s \textit{maximum pairwise domination} is equal to $\max_{b \neq a} |\{\pi_i | a \succ_{\pi} b\}|$. } \textsc{Kemeny-Young} selects the candidate ranked first in the ranking with the minimal sum of \textit{Kendall tau} distances from voters' rankings.\footnote{The Kendall tau distance between rankings $\ranking,\ranking'$ is the total number of swaps required to transform $\ranking$ into $\ranking'$.}  \textsc{Copeland} selects the alternative with the most points, giving $a$ 1 point for each candidate $b$ it pairwise dominates, and 1/2 point for each $b$ with which it pairwise ties. Finally, in our analysis, we will often study a general class of voting rules called \textit{hyperplane rules}. This class is known to be equivalent to \textit{generalized scoring rules} \cite{xia2008generalized} and encompasses essentially every standard voting rule, including all those listed above.  
\begin{definition}[Hyperplane rules \cite{mossel2013smooth}] \label{def:hyperplane_rules} Note that given a set of $\ell$ affine-hyperplanes, these hyperplanes partition the space of histograms into at most $3^\ell$ regions, as every point is either on a hyperplane or on one of two sides. We say that $R$ is a \emph{hyperplane rule} if there exists a finite set of affine hyperplanes $H_1, \ldots , H_\ell$ such that $R$ is constant on each such region.
\end{definition}
We will sometimes subdivide this class of rules further: \textit{decisive} hyperplane rules are those which output a single winner on profiles that are not on any hyperplane; \textit{non-decisive} hyperplane rules are all others.\\[-0.5em]

\noindent \textbf{Axioms.} Formally, an \textit{axiom} is a function $A : \mathcal{R} \to (\profiles \to \set{\textsf{True},\textsf{False}})$, mapping a voting rule $R$ to another mapping describing whether $A$ is satisfied by $R$ on any given profile. We can think of $A(R)$ as representing the true/false statement ``$R$ is consistent with $A$ on $\profile$''. 
We will study several standard axioms, defined formally in \Cref{app:axioms} and described colloquially below.

$R$ satisfies \textsc{Resolvability} on $\profile$ if it selects a single winner. $R$ satisfies \textsc{Condorcet Consistency} (abbreviated as \textsc{Condorcet}) on $\profile$ if it selects the \textit{Condorcet winner} (the candidate that pairwise dominates all other candidates), or by default if $\profile$ has no Condorcet winner. $R$ satisfies \textsc{Majority} on $\profile$ if it selects the \textit{majority winner} (the candidate ranked first by a majority of voters), or by default if $\profile$ does not have a majority winner. $R$ satisfies \textsc{Consistency} on $\profile$ if there is no partition of $\profile$ into subprofiles such that a unique candidate $b$, which is \textit{not} the winner in $\profile$, is chosen as the winner on all subprofiles in that partition. $R$ satisfies \textsc{Independence of Irrelevant Alternatives (IIA)} on $\profile$ if the winner, $a$, cannot be made to lose to $b$ by adjusting votes in a way that does not change the relative positions of $a$ and $b$.
%

We also use $\rho(n)$\textsc{-Group-Stability}, which colloquially requires that the outcome of voting rules be \textit{stable} to a change in the behavior of up to $\rho(n)$ voters. This is essentially framing \emph{margin of victory} as a binary condition.
\begin{definition}[$\rho(n)$\textsc{-Group-Stability}] \label{def:group_stability}
For a given rule $R$, $\rho(n)$\textsc{-Group-Stability}($R$) is satisfied if, for every pair of profiles $\profile,\profile'$ that differ on at most $\rho(n)$ of voters' rankings, $R(\profile) = R(\profile')$. 
\end{definition}
This axiom implies strong, $\rho(n)$-parameterized group-level versions of three axioms of common interest: $\rho(n)$-\textsc{Group-Strategyproofness}\emdash no group of up to $\rho(n)$ voters can strategically misreport their preferences in concert and cause $R$ to output an alternative they all weakly prefer, and at least one of them strongly prefers; $\rho(n)$-\textsc{Group-Participation}\emdash no group of up to $\rho(n)$ voters can leave the election and cause $R$ to output an alternative they all weakly prefer, and at least one of them strongly prefers);  and $\rho(n)$-\textsc{Group-Monotonicity}\emdash no group of up to $\rho(n)$ voters can weakly decrease the position of an alternative $c$, which is not currently a winner by $R$, in their rankings, and cause $c$ to become a winner by $R$.\footnote{The fact that $\rho(n)$-\textsc{Group-Stability} implies the first and third of these axioms is by definition. For the second, $\rho(n)$-\textsc{Group-Stability} technically implies $\frac{\rho(n)}{1-\rho(n)}$-\textsc{Group-Participation}, since this axiom involves $\rho(n)$ voters leaving the electorate; this does not change any of the asymptotic results, and we handle it in our proofs.} We formally define these axioms in \Cref{app:axioms}.

\subsection{Smoothed model of profiles} \label{sec:noisemodel}
\textbf{Noise distributions over rankings.} A \textit{noise distribution} is effectively a distribution over rankings; formally, it is a distribution over permutations $\sigma: [m] \to [m]$. When we ``apply noise'' to a ranking $\pi$ via a noise distribution $\noisemodel_\phi$ (with parameter $\phi$ to be defined later), we are drawing a random permutation $\sigma \sim \noisemodel_\phi$, and then permuting $\ranking$ according to $\sigma$.  
Abusing notation slightly, we represent the distribution over rankings induced by perturbing $\ranking$ with noise model $\noisemodel_\phi$ as $\noisemodel_\phi(\ranking)$.\footnote{More formally, $\noisemodel_\phi(\ranking)=\ranking + \sigma$ with $\sigma \sim \noisemodel_{\phi}$. Here, the $+$ operator represents composition: if $\sigma(i) = j$, then the $i$-th ranked candidate in the perturbed ranking will be $\ranking(j)$.} 

In this paper, we consider the class of noise distributions $\boldsymbol{\noisemodel}$,  encompassing any distribution over rankings satisfying the minimal regularity conditions in Assumptions 1 and 2 below.  All $\noisemodel_\phi \in \boldsymbol{\noisemodel}$ are parameterized by a dispersion parameter $\phi \in [0,1]$, which captures the ``level'' of noise applied by $\noisemodel_\phi$. We will use $\phi$ to implement multiple different such measures throughout our results. Assumptions 1 and 2 impose the minimal requirements to ensure that $\phi$ reasonably measures the amount of noise, and that the noise distribution is practical to work with. As we implement specific uses of $\phi$, we will impose additional assumptions as needed in later sections. 

\begin{assumption}[Extremal values] \label{ass:extremal}
The distribution $\noisemodel_0$ is the point mass on the identity (so that $\phi = 0$ corresponds to no noise added). The distribution $\noisemodel_1$
is uniform over all permutations (so that $\phi = 1$ corresponds maximum noise). Note that this implies that for all profiles $\profile \in \profiles$, $\noisemodel_1(\profile)$ is equivalent to the impartial culture model (see, e.g.,~\cite{eugeciouglu2013impartial}).
\end{assumption}

\begin{assumption}[Continuity] \label{ass:continuity}
For all $\sigma$, the probability $\noisemodel_\phi$ places on $\sigma$ is continuous in $\phi$.
\end{assumption}

\noindent \textbf{Sampling noisy profiles.} Applying noise via $\noisemodel_\phi$ to an entire \textit{profile} means applying it to every ranking within it \textit{independently} (an assumption also made in the semi-random model \cite{xia2020smoothed}).
Formally, if $\profile$ is our starting profile, then our noisy profile is drawn from the distribution $\prod_{i = 1}^n \rankingdistrankingi$, where each $\rankingdistrankingi$ is independent. The resulting noisy profile is denoted $\noisemodel_\phi(\profile)$.  
We will treat $\noisemodel_\phi$, $\noisemodel_\phi(\ranking)$, and $\noisemodel_\phi(\profile)$ as distributions and random variables interchangeably.

Since we will work with profiles in histogram form, so we will usually reason about distributions over rankings and profiles \textit{projected into histograms space}. We express the distribution over ranking histograms induced by noise distribution $\noisemodel_\phi(\ranking)$ as $\prop(\noisemodel_\phi(\ranking))$. Where the former is a distribution over rankings, the latter is a distribution over basis vectors. Then, the corresponding noise distribution over profile histograms is, naturally, $
    \propdistprof = \nicefrac{1}{n} \sum_{i = 1}^n \propdistrankingi.$

\subsection{Smoothed-satisfaction and smoothed-violation of axioms} \label{sec:smoothed}
First, we formally define worst-case notions of axiom satisfaction and violation. Recall that $A(R) : \profiles \to \{\textsf{True,False}\}$ intakes a given profile and outputs whether rule $R$ satisfies axiom $A$ on that profile. Then, we say that $\profile$ is a \textit{counterexample} to $A(R)$ iff $(A(R))(\profile) = \textsf{False}$. Let $\profiles^{\neg(A(R))}$ be the set of all profiles that are counterexamples to $A(R)$.
Then, in the worst-case sense, $R$ \textit{satisfies} $A$ if $\profiles^{\neg(A(R))} = \emptyset$ (i.e., no counterexample exists); otherwise, $R$ \textit{violates} $A$. 

Now, we define what it means for $R$ to satisfy or violate $A$ in the smoothed model. Conceptually, $R$ \textit{smoothed-satisfies} $A$ if the probability that $R$ satisfies $A$, after applying a noise distribution, converges to 1 as $n$ grows large. In contrast, $R$ \textit{smoothed-violates} $A$ if probability of $R$ satisfying $A$ being satisfied converges to 0 as $n$ grows large. Note that worst-case satisfaction implies smoothed-satisfaction, and smoothed-violation implies violation.

\begin{definition}[smoothed-satisfied] \label{def:brittle}
Voting rule $R$ $\noisemodel$-\textit{smoothed-satisfies} axiom $A$ at a rate $f(n,\phi)$ if, for all $n \in \mathbb{Z}^+$ and $\phi \in (0,1]$, 
    \[
        \sup_{\phi' \in [\phi, 1]} \sup_{\profile \in \profiles_n} \Pr\left[\noisemodel_{\phi'}(\profile) \in \profiles^{\neg A(R)}\right] \le f(n,\phi).
    \]

\end{definition}
\begin{definition}[smoothed-violated] \label{def:robust} A voting rule $R$ $\noisemodel$-\textit{smoothed-violates} axiom $A$ at a rate of $f(n)$ if there exists $\phi \in (0, 1]$ and a profile $\profile$ of size $n$ such that for all $z \in \mathbb{Z}^+$, 
\[
        \inf_{\phi' \in [0, \phi]} \Pr\left[\noisemodel_{\phi'}(z\profile) \in\profiles^{\neg A(R)}\right] \ge 1- f(zn).
\]

\end{definition}

Per \Cref{def:brittle}, convergence to smoothed-satisfaction occurs eventually for \textit{all} $\phi \in (0,1]$, although the rate depends on $\phi$.
When we show smoothed-violation, in contrast, we are saying there exists a constant amount of noise $\phi$ such that this amount, or any less, will not be enough to ensure satisfaction of $A$ by $R$ as $n$ grows large. Note the gap between these two definitions: smoothed-violated is not the negation of smoothed-satisfied, and a claim could therefore be ``in-between'' these definitions, satisfying neither. This will not end up being the case for any of the voting rules or criteria we study.

\section{Patterns across smoothed-violated, smoothed-satisfied axioms} \label{sec:results1}

\subsection{Condorcet, Majority, Consistency, 
IIA} \label{sec:negative-results}
This section is dedicated to axioms about which we prove negative results. As summarized in \Cref{tab:classification}, we find that for all the voting rules we study, smoothed noise is insufficient to prevent existing violations of any of these axioms. Across these axioms, the reason for this insufficiency is the same: as formalized in our generic sufficient condition in \Cref{thm:violation-sufficient}, the histogram space contains contiguous regions of counterexamples.

\begin{table}[ht]
    \centering
    \begin{tabular}{r||cccc}
        \multicolumn{1}{l}{} & \multicolumn{4}{c}{\textbf{Axioms}}\\
         \textbf{Voting Rules} &   \textsc{Condorcet}  &  \textsc{Majority$^*$} &   \textsc{Consistency$^*$} & \textsc{IIA$^*$}\\
         \hline \hline
        \textsc{Plurality} &  \cellcolor{red!10} s-violated & \textcolor{gray}{satisfied} &   \textcolor{gray}{satisfied} &  \cellcolor{red!10} s-violated \\
        PSRs $\setminus$ \textsc{Plurality} &  \cellcolor{red!10} s-violated & \cellcolor{red!10} s-violated &  \textcolor{gray}{satisfied} &  \cellcolor{red!10} s-violated \\
        \textsc{Minimax} &  \textcolor{gray}{satisfied} &  \textcolor{gray}{satisfied} & \cellcolor{red!10}  s-violated &  \cellcolor{red!10} s-violated \\
        \textsc{Kemeny-Young}  &   \textcolor{gray}{satisfied} &   \textcolor{gray}{satisfied} &   \cellcolor{red!10}  s-violated &  \cellcolor{red!10} s-violated \\
        \textsc{Copeland}  &    \textcolor{gray}{satisfied} &   \textcolor{gray}{satisfied} &     \cellcolor{red!10} s-violated & \cellcolor{red!10} s-violated  \\
    \end{tabular}
    \caption{Summary of results stated in \Cref{prop:CSRS_axioms}. s-violated = smoothed-violated.}
    \label{tab:classification}
\end{table}

\begin{theorem} \label{prop:CSRS_axioms}
For all  $A \in \{\textsc{Condorcet, Majority, Consistency, IIA}\}$ and all $R \in \text{PSRs} \cup \{\textsc{Minimax},\textsc{Kemeny-Young},\textsc{Copeland}\}$, if $R$ violates $A$, then $R$ smoothed-violates $A$.
\end{theorem}
The proof of this theorem is found in \Cref{app:CSRS_axioms}; we explain the intuition here. The core idea is that for small enough $\phi$, $\propdistprof$ concentrates at an exponential rate very near the starting histogram $\proportions^{\profile}$ (\Cref{lem:starting-concetration}). This lemma, proven in \Cref{app:starting-concentration}, follows from the fact that noise is applied independently across rankings, allowing the application of a simple Hoeffding bound.
\begin{lemma}\label{lem:starting-concetration}
    Let $\noisemodel$ be a noise model. For all $\e > 0$, there exists a $\phi \in (0, 1]$ such that for all $\phi'  \in [0, \phi]$ and profiles $\profile \in \profiles_n$ on $n$ voters,
    $$
    \Pr\left[\left\| \prop(\noisemodel_{\phi'}(\profile)) - \proportions^{\profile} \right\|_1  < \e\right] > 1 - \exp \left(\e^2 n/2 \right).
    $$
\end{lemma}
  
 From this concentration follows a general sufficient condition for smoothed-violation (\Cref{thm:violation-sufficient}): that there exists a counterexample $\profile$ such that $\proportions^{\profile}$ is contained within a ball of counterexamples. Then, as the noise distribution concentrates around $\proportions^{\profile}$, most of its probability mass is contained in the ball, and the probability of a counterexample converges to 1. 

\begin{lemma}\label{thm:violation-sufficient}
    Fix a noise model $\noisemodel$, axiom $A$, and rule $R$. Suppose there exists a profile $\profile$ and radius $r > 0$ such that
    for all profiles $\profile' \in \profiles$ satisfying
    (1) $|\profile'| = z|\profile|$ for some $z \in \mathbb{Z}^+$ and
        (2) $\|\proportions^{\profile'} - \proportions^\profile\|_1 < r$, it is the case that 
    $\profile' \in \profiles^{\neg A(R)}$. Then,
    $R$ \textbf{smoothed-violates} $A$ at a rate of
    $$f(n) = \exp \left(-r^2 n/2 \right).$$
\end{lemma}
\begin{proof}
    Fix a noise model $\noisemodel$, an axiom $A$, and a voting rule $R$. Fix a profile $\profile$ and radius $r > 0$ satisfying the preconditions of the theorem.
Choose $\e = r$ and let $\phi$ be the one from \Cref{lem:starting-concetration} corresponding to $\e$ and $\noisemodel$. Fix an arbitrary $z \in \mathbb{Z}^+$ and $\phi' \in [0, \phi]$. Notice that $\proportions^{\profile} = \proportions^{z \profile}$. Hence, \Cref{lem:starting-concetration} guarantees that the post-noise histogram lies in an $r$-radius ball around the original histogram $\proportions^\profile$ with high probability\emdash that is, $\prop(\noisemodel_{\phi'}(z\profile)) \in B^{L_1}_r(\proportions^{\profile})$ with probability at least $1-f(z|\profile|) = 1 - \exp \left(-r^2 z|\profile|/2 \right)$. Note also that every profile in the support of $\noisemodel_{\phi'}(z\profile)$ is guaranteed to have $z|\profile|$ voters. These two facts, taken with both preconditions of the theorem, imply that the post-noise profile histogram is a counterexample with high probability\emdash that is, $\noisemodel_{\phi'}(z\profile) \in \profiles^{\neg A(R)}$ with probability at least $1-f(z|\profile|)$. 
It then follows, by the definition of smoothed violation (\Cref{def:robust}), that $R$ smoothed-violates $A$ at a rate of $f(z|\profile|)$, as needed.
\end{proof}

Finally, we conclude \Cref{prop:CSRS_axioms} by finding counterexamples for each $A,R$ that exist within balls of other counterexamples, and then applying \Cref{thm:violation-sufficient}. Across rules and axioms, simple counterexamples suffice, supporting the intuition that the insufficiency of smoothed noise is not a quirk of these rules and axioms. Rather, smoothed noise may be insufficient across many interpretable rules and axioms, because\emdash as a natural consequence of their interpretability\emdash they behave similarly on similar profiles.

\subsection{\textsc{Resolvability}, \textsc{Strategyproofness}, \textsc{Participation,} and \textsc{Monotonicity}} \label{sec:positive-results}
This section is dedicated to axioms about which we prove (mostly) positive results, summarized in \Cref{tab:classification2}. Across these axioms, our proofs use a common property that makes smoothed-noise \textit{sufficient} for circumventing impossibilities: that across voting rules, their counterexamples are restricted to regions on or near a limited number of hyperplanes.

\begin{table}[h!]
    \centering
    \begin{tabular}{lr||cc}
        \multicolumn{1}{l}{} & & \multicolumn{2}{c}{\textbf{Axioms}}\\
          & \textbf{Voting Rules}&  \textsc{Resolvability}  &  \textsc{$o(\sqrt{n})$-Group Stability$^*$}\\
         \hline \hline
         & \textsc{Plurality} & \cellcolor{blue!10}  s-satisfied & \cellcolor{blue!10}  s-satisfied \\
        \textcolor{gray}{(Decisive} & PSRs $\setminus$ \textsc{Plurality} &   \cellcolor{blue!10} s-satisfied &  \cellcolor{blue!10} s-satisfied\\
        \textcolor{gray}{hyperplane} & \textsc{Minimax} & \cellcolor{blue!10}  s-satisfied &  \cellcolor{blue!10} s-satisfied \\
        \textcolor{gray}{rules)} & \textsc{Kemeny-Young}  &   \cellcolor{blue!10} s-satisfied & \cellcolor{blue!10} s-satisfied\\
        \hline
        & \textsc{Copeland}  &   \cellcolor{red!10} s-violated & \cellcolor{blue!10} s-satisfied \\
    \end{tabular}
    \caption{Summary of results stated in \Cref{prop:resolvability} and \Cref{prop:group-stability}. s-satisfied = smoothed-satisfied. Positive results for $o(\sqrt{n})$-\textsc{Group-Stability} directly imply positive results for $o(\sqrt{n})$-\textsc{Group-Strategyproofness}, $o(\sqrt{n})$-\textsc{Group-Participation}, and $o(\sqrt{n})$-\textsc{Group-Monotonicity.}}
    \label{tab:classification2}
\end{table}

As are upper bounds in Xia's semi-random model \cite{xia2020smoothed}, our upper bounds here will be parameterized by $\minprob(\noisemodel_\phi) := \min_{\sigma} \Pr[\noisemodel_\phi = \sigma]$, the smallest probability $\noisemodel_\phi$ assigns to any permutation. This parameterization motivates two additional assumptions: \textsc{\Cref{ass:positivity}} ensures that $1/\minprob(\noisemodel_\phi)$ is well-defined, and \textsc{\Cref{ass:monotonicity}} describes how $\minprob$ implements $\phi$ as a measurement of the noisiness of $\mathcal{S}_\phi$. 
\begin{assumption}[Positivity] \label{ass:positivity}
For all $\phi \in (0,1]$, $\minprob(\noisemodel_\phi) > 0$. That is, for any nonzero amount of noise, the resulting distribution assigns positive probability to all permutations. 
\end{assumption}

\begin{assumption}[Weak Monotonicity] The value $\minprob(\noisemodel_\phi)$ is non-decreasing in $\phi$.\footnote{We note that our results don't centrally depend on \Cref{ass:monotonicity}\emdash \Cref{ass:continuity,ass:positivity} are sufficient to give the same high-level results. We include \Cref{ass:monotonicity} because it is not prohibitive, it simplifies the exposition, and allows us to give more useful parameterized bounds.} \label{ass:monotonicity}
\end{assumption}

First, \Cref{prop:resolvability} shows that \textsc{Resolvability} is smoothed-satisfied for all decisive hyperplane rules. Notably, these rules exclude the known rule \textsc{Copeland}, due to its lack of sensitivity to changes in rankings in regions surrounding ties.
\begin{theorem}\label{prop:resolvability}
    All decisive hyperplane rules smoothed-satisfy \textsc{Resolvability} at a rate\\ $O_m(\nicefrac{1}{\sqrt{n} \cdot \minprob(\noisemodel_{\phi})^{3/2}})$. All non-decisive hyperplane rules smoothed-violate \textsc{Resolvability}. 
\end{theorem}
The formal proof is found in \Cref{app:resolvability}. 
This result corely relies on the fact that, essentially regardless of the noise distribution over \textit{rankings} $\noisemodel_\phi$, the resulting noise distribution over \textit{entire profile histograms} $\propdistprof$ must converge uniformly\footnote{``Uniform'' (over profiles)  convergence means that for all profiles $\profile$ of any fixed $n$, $\propdistprof$ is at most $O(\nicefrac{1}{\sqrt{n}})$ ``distance away'' from the the Gaussian distribution with expectation and variance corresponding to that of $\propdistprof$.} to a multi-dimensional Gaussian distribution at a rate of $O(1/\sqrt{n})$  (\Cref{lem:normality}):

\begin{lemma}\label{lem:normality}
        Let $\noisemodel$ be a noise model, $\phi \in [0,1]$ a parameter, and $\profile \in \profiles_n$ a profile on $n$ voters. Then, for all convex sets $X \subseteq \R^{m! - 1}$,
        \vspace{-0.5em}
        $$\left|\Pr \big[ \propdistprof \in X\big] - \Pr \big[\mathcal{N}\left(\,\E[\propdistprof],\, \Cov[\propdistprof]\,\right) \in X\big]\right| \le \frac{ O((m!)^{7/4})}{\sqrt{n} \cdot \minprob(\noisemodel_\phi)^{3/2}}.$$
\end{lemma}
\noindent Intuitively, $\propdistprof$ converges to a Gaussian because it is akin to the sum of independent indicators (literally, independently-drawn rankings represented in histogram space as basis vectors). We prove this convergence in \Cref{app:normality} using a general form of the Berry Esseen bound \cite{bentkus2005lyapunov}. 

This convergence allows us to prove a general sufficient condition for the smoothed-satisfaction: that the set of counterexamples are contained with a measure-zero subset of histograms (\Cref{thm:brittle_convex}).

\begin{lemma} \label{thm:brittle_convex}
Fix a noise model $\noisemodel$, voting rule $R$, and axiom $A$. If there exists some set $X$ such that (1) $X= \bigcup_{j = 1}^\ell X_j$ where each $X_j \subseteq \Delta_{\proportions}$ is convex, (2) $\prop\left(\profiles^{\neg A(R)}\right) \subseteq X$, and (3) $X$ is measure zero (noting that $X$ is necessarily measurable), then $R$ \textbf{smoothed-satisfies} $A$ at a rate of
    $$f(n,\phi)= \frac{\ell \cdot O((m!)^{7/4})}{\sqrt{n} \cdot \minprob(\noisemodel_{\phi})^{3/2}}.$$
\end{lemma}
\noindent The formal proof is found in \Cref{app:brittle_convex}, and uses that the Gaussian places zero probability mass over measure-zero regions of its support. This concludes our analysis of \textsc{Resolvability}.\\[-0.5em]

Now, we show that $\rho(n)$-\textsc{Group-Stability} (\Cref{def:group_stability}) is smoothed-satisfied by all hyperplane rules for $\rho(n) \in o(\sqrt{n})$. Because $\rho(n)$-\textsc{Group-Stability} implies $\rho(n)$-\textsc{Group-Strategyproofness}, $\rho(n)$-\textsc{Group-Participation}, and $\rho(n)$-\textsc{Group-Monotonicity}, these axioms must also be smoothed-satisfied by all hyperplane rules. 

\begin{theorem} $o(\sqrt{n})$-\textsc{group-stability} is smoothed-satisfied by all hyperplane rules at a rate of $O_m(\nicefrac{1}{\sqrt{n} \cdot \minprob(S_{\phi})^{3/2}}) + o(1)$. \label{prop:group-stability} 
\end{theorem} 
\noindent We will now prove \Cref{prop:group-stability} via the following anti-concentration lemma, which states that for any $\profile$, $\propdistprof$ places limited probability mass within distance $\delta(n)$ of any specific hyperplane as $n$ grows large, so long as $\delta(n)$ is decreasing sufficiently quickly in $n$. 
\begin{lemma} \label{thm:generalized_brittleness}
Let $\mathcal{G}$ be the set of all hyperplanes in $\R^{m! - 1}$.
For all noise models $\noisemodel$, parameters $\phi \in [0, 1]$, and $\delta(n) \in o(1/\sqrt{n})$, we have the following, where $d$ is the $L_1$ distance.
$$\sup_{G \in \mathcal{G}}\sup_{\phi' \in [\phi, 1]}\sup_{\profile \in \profiles_n}\Pr\left[d(\prop(\noisemodel_{\phi'}(\profile)), G) \le \delta(n)\right] \in O\left(\frac{\delta(n)\sqrt{n}}{\sqrt{\minprob(S_\phi})}  + \frac{1}{\minprob(\noisemodel_\phi)^{3/2} \sqrt{n} }\right) \in o(1).$$
\end{lemma}
This lemma, proven in \Cref{app:gen_brittleness}, shows that even if $\mathbb{E}[\prop(\noisemodel_\phi(\profile))]$ falls within $\delta(n)$ distance of the hyperplane (we can think of this as it falling within a ``thick'' hyperplane), the width of this thick hyperplane is shrinking faster than the distribution over histograms concentrates as $n$ grows large. To make the convergence rate in \Cref{thm:generalized_brittleness} $o(1)$, $\delta(n)$ must be in $o(1/\sqrt{n})$.

To apply this lemma to show the smoothed-satisfaction of $\rho(n)$\textsc{-Group-Stability}, first observe that for a group of size $\rho(n)$ to be able to impact the outcome of the election in $\profile$ (i.e., for $\profile$ to be a counterexample), that group must be \textit{pivotal} in $\profile$\emdash that is, $\proportions^{\profile}$ must lie within some $\rho(n)$-dependent distance from a profile on which the winner changes. Because the set of such profiles are defined by finitely-many hyperplanes (by definition of hyperplane rules), counterexamples are then restricted to ``thick'' hyperplanes. To apply \Cref{thm:generalized_brittleness} for each such hyperplane, we need their width to be $o(1/\sqrt{n})$ in histogram space, corresponding to a coalition of size $\rho(n) \in o(\sqrt{n})$. Then, we conclude \Cref{prop:group-stability} by simply union bounding over the finite number of hyperplanes referred to in \Cref{def:hyperplane_rules}. \qed

\section{Beyond dependence on the minimum probability}\label{sec:beyond-minprob}
All convergence rates to smoothed-satisfaction proven so far\emdash both in \Cref{sec:positive-results} and in the related work \cite{xia2020smoothed}\emdash depend on $1/\sqrt{n \cdot \minprob^3}$. As a result, when $\minprob$ is very small, extremely large $n$ is required to get reasonably low probability of an axiom violation: examining the relative $n$ and $\minprob$ dependency above, if we decrease $\minprob$ by factor $\ell$, we need to increase the number of voters $n$ by a factor of $\ell^3$ in order to recover the same probability bound.\footnote{One may wonder if different techniques could potentially improve this dependence. Although we do not have a tight cubic lower bound and the exact bound should depend on the exact voting rule/axiom, we can at least get a linear one. Indeed, consider the noise distribution that switches to any ranking other than the starting one with probability $\varepsilon$, but stays on the starting one with probability $1 - (m! - 1)\varepsilon$. The minimum probability here is $\varepsilon$, but we need $n \in \omega(\varepsilon / m!)$ to ensure that the profile changes at all with high probability, a necessary condition to smoothed-satisfy any axiom that fails on even a single profile.}

Motivated by the lack of good bounds for the important class of noise models with small $\minprob$, we now pursue the first $\minprob$-independent bounds on convergence rates. For concreteness, we specifically pursue these bounds within the well-established \textit{Mallows model}, as defined below; however, as we will illustrate in detail throughout this section, our results rely only weakly on the properties of this model, and should apply to more general models as well. 

\begin{definition}[Mallows noise model \cite{mallows1957non}]\label{def:mallows}
Let $d : \rankings \times \rankings \to \mathbb{N}$ be the Kendall tau distance, and let $\phi \in [0,1]$. Then, the Mallows model $\noisemodel^{Mallows}_\phi$ is defined as follows, where $Z = \sum_{\ranking' \in \rankings} \phi^{d(\ranking',\ranking)}$ is a normalizing term.\vspace{-0.5em}
\[\Pr\left[\noisemodel^{Mallows}_{\phi}(\ranking) = \ranking'\right] = \frac{1}{Z} \phi^{d(\ranking,\ranking')}.\]
\end{definition}
The Mallows model is an attractive case study for proving $\minprob$-independent bounds for two reasons. First, it precisely captures the intuition motivating this analysis: that long-range swaps should be rare under less noise, making $\minprob$ low. Longer range swaps are \textit{so} rare in this model, in fact, that $\minprob(\noisemodel^{Mallows}_\phi)$ approaches zero exponentially fast as $m$ grows: the maximum Kendall tau distance between rankings is $\binom{m}{2}$, so $\minprob(\noisemodel^{Mallows}_\phi) \in \phi^{\Omega(m^2)}$. This motivates our second reason: that despite its importance in social choice, existing convergence rates grow poorer 
 at an exponential rate for Mallows noise as $m$ gets large and/or $\phi$ gets small.

Within the Mallows model, we characterize the precise $m, \phi, n$-dependent rates at which \textsc{Resolvability} and $o(\sqrt{n})$-\textsc{Strategy-proofness} are smoothed-satisfied by four diverse voting rules: \textsc{Plurality}, \textsc{Borda}, \textsc{Veto}, and \textsc{Minimax}. Recall that these voting rules
are already known to smoothed-satisfy both axioms by our results in \Cref{sec:positive-results}. The key difference in this analysis is that, while before $n$ was being treated as the only variable (with $m$ and the noise level treated as constants), we now consider more closely the convergence depends on $m$ and $\phi$. In particular, we are interested in how large $n$ must be (as a function of $m$ and $\phi$) for the rate to be $o(1)$ (i.e., satisfaction occurring with high probability). 
We summarize our results in \Cref{tab:classification4}, framed to directly answer this question. The formal statements and proofs are below, in \Cref{sec:mallows-resolve}.

\begin{table}[h!]
    \centering
    \begin{tabular}{r||ccc}
        & \multicolumn{3}{c}{\textbf{Axioms}}\\
           \textbf{Voting Rules}&  \textsc{Resolvability}  &  \textsc{$n^p$-Group-Strategyproof}& \\
         \hline \hline \\[-0.9em]
         \textsc{Plurality} & $\omega(\nicefrac{m^5}{\phi})$ is sufficient & $\omega((\nicefrac{m^5}{\phi})^{\nicefrac{1}{(1 - p)}})$ is sufficient& (Prop.~\ref{lem:plurality-mallows}) \\[0.25em]
        \textsc{Borda} &  $\omega\left(\nicefrac{m^6}{ \phi}\right)$ is sufficient & $\omega((\nicefrac{m^8}{\phi})^{\nicefrac{1}{(1 - p)}})$ is sufficient & (Prop.~\ref{lem:borda-mallows}) \\[0.25em]
                \hline \\[-0.9em]
        \textsc{Veto} &  $\Omega(\nicefrac{1}{\phi^{m - 2}})$ is necessary &  $\Omega(\nicefrac{1}{\phi^{m - 2}})$ is necessary & (Prop.~\ref{lem:veto-mallows})\\
        \textsc{Minimax} &  $\Omega\left(\nicefrac{1}{m\phi^{\floor{m/2}}}\right)$ is necessary & $\Omega\left(\nicefrac{1}{m\phi^{\floor{m/4}}}\right)$ is necessary& (Prop.~\ref{lem:minimax-mallows})
    \end{tabular}
    \caption{}
    \label{tab:classification4}
\end{table}
What is striking in these results is a clear separation between voting rules, which was not visible in our $\minprob$-parameterized bounds. The convergence rates of \textsc{Plurality} and \textsc{Borda} do not get dramatically worse as $\phi$ (and thus $\minprob$) gets small; put another way, as $\phi$ scales down, $n$ must scale up proportionally to maintain roughly the same probability of satisfaction. In contrast, \textsc{Minimax} and \textsc{Veto} require exponentially (in $m$) large $n$.\footnote{To put this in perspective, suppose $m = 6$ and we decrease $\phi$ from $1/5$ to $1/10$. To maintain a similar probability of violating \textsc{Resolvability}, if we are using \textsc{Plurality} or \textsc{Borda} it is sufficient to double the number of voters; if we are using \textsc{Veto}, one needs at least $2^4 = 32$ times as many voters, and for \textsc{Minimax}, one needs at least $2^3 = 8$ times as many voters. This gap only gets steeper as $m$ gets larger.} 

We can, as before, distill a pattern explaining this gap: the voting rules that achieve the best rates (\textsc{Plurality, Borda} are those whose outcomes are more sensitive to local swaps across the support (or, in critical areas of the support, in the case of \textsc{Plurality}). The outcomes of \textsc{Veto} and \textsc{Maximin}, in contrast, are fairly insensitive to local swaps, allowing us to show that local swaps are not enough to overcome impossibilities. While this pattern is perhaps unsurprising in retrospect, it may be important in informing the choice of voting rules.

\subsection{Formal statements and proofs} \label{sec:mallows-resolve}
We include in the body the proofs for one upper bound (\textsc{Plurality}) and one lower bound (\textsc{Veto}), and defer the proofs for \textsc{Borda} and \textsc{Minimax} to \Cref{app:borda-mallows,app:minimax-mallows}, respectively. Our arguments will rely on only very weak properties of the Mallows model, essentially requiring just that the noise rarely induces swaps between distantly-ranked candidates. To emphasize the kinds of noise models to which our arguments can generalize, we now recap the precise properties of the noise model required for the proof corresponding to each voting rule. For \textsc{Plurality}, the key property of the noise distribution we use is that no alternative is ranked first post-noise with probability near $1$. For \textsc{Borda}, the key property is that there is no $\ell$ for which two candidates will be exactly $\ell$ positions apart post-noise with probability near $1$. For \textsc{Veto}, the key property is that there is an exponentially small probability of a given voter moving one of their top two candidates to last place. For \textsc{Minimax}, the key property is that it is unlikely for two candidates a distance of $m/2$ apart to swap. Before presenting our results, we establish a few useful properties of Mallows noise.

\begin{lemma}[First- and last-place probability \cite{awasthi2014learning}]\label{lem:first-last}
    From starting ranking $\pi$, the probability that $\pi(j)$ is ranked first in a ranking drawn from
    $\noisemodel^{Mallows}_{\phi}(\ranking)$ is proportional to $\phi^j$, i.e., $\phi^j/\sum_{j' = 1}^m \phi^{j'}$. Symmetrically, the probability $\pi(j)$ is ranked last is proportional to $\phi^{m - j}$. 
\end{lemma}

From \cite{mallows1957non}, if two candidates $i,j$ are $k$ positions apart in $\pi$ (i.e., $\pi(i)$ and $\pi(j)$ with $i - j = k$), the probability that they retain their relative order post-Mallows noise is increasing in $k$, and is always at least $1/2$. This was refined by \cite{mallowsprobbound} to an exact value of their swap probability:
\begin{lemma}[swap probability \cite{mallowsprobbound}] For candidates $i,j$ such that $\pi(i)$ and $\pi(j)$ with $i - j = k$, their probability of swapping post-Mallows noise is
\[q(k) := \frac{1}{1-\phi^{k+1}}\left(1 - \frac{(1-\phi)k\phi^k}{1 - \phi^k}\right).\]
\end{lemma}
We will often parameterize our bounds by the probability of the opposite event\emdash that $i$ and $j$ at distance $k$ \textit{do} swap, which we denote by $\bar{q}(k) := 1 - q(k)$. We will use the following upper bounds on $\bar{q}(k)$, derived in \Cref{app:mallows-ubs}. 
\begin{lemma} \label{lem:mallows-ubs}
For all $k \in [m]$ and $\phi \in [0,1]$, $\bar{q}(k) \leq \min\left\{k\phi^k, \phi^{k/2}, \frac{\phi}{1 + \phi}\right\}.$
\end{lemma}

\begin{proposition}[\textsc{Plurality}] \label{lem:plurality-mallows}
    \textsc{Plurality} smoothed-satisfies \textsc{resolvability} at a rate of at most \vspace{-0.5em} 
    \[O\left(m^{5/2}/\sqrt{n \phi} + m\exp\left(-n / 6m\right)\right)\]
    and, as long as $\rho(n) \le \frac{n}{6m}$, smoothed-satisfies $\rho(n)$\textsc{-group-stability} at a rate of
    \[
        O\left(\rho(n) m^{5/2}/\sqrt{n \phi} + m\exp\left(-n / 9m\right)\right).
    \]
\end{proposition}
\begin{proof}
We begin with resolvability and show how to extend it to $\rho(n)$\textsc{-Group-Stability} after. Fix a starting profile $\profile$.
For a candidate $c$, let $X^c_i$ be the indicator that voter $i$ ranks $c$ first, and let $p^c_i = \Pr[X^c_i = 1]$. Then, let $S^c = \sum_i X^c_i$ be the random variable representing plurality score of $c$ post-noise. It follows that $\mathbb{E}[S^c] = \sum_{i \in [n]}p_i^c.$ We will partition the candidates into two sets based on these expectations: $L$ (for ``low'') is defined as $L = \{c | \mathbb{E}[S^c] < \frac{n}{2m}\}$, and $H$ (for ``high'') is defined as $H =\{c | \mathbb{E}[S^c] \ge \frac{n}{2m}\}$. We will first show \textit{Claim 1:} with high probability, the winner will be from $H$. 
Then, in \textit{Claim 2}, we will show that the probability any two candidates in $H$ have the same plurality score is small. Union bounding over these two events will upper bound the probability of an unresolvable outcome, i.e., a tie in plurality scores.\\[-0.5em]

\noindent \textit{Proof of Claim 1.} Fix a candidate $a \in L$.  Note that a necessary condition for $a$ to be a plurality winner is for their score to be at least $n/m$. Standard Chernoff bounds says that $\Pr[S^a \ge (1 + \delta)\mu] \le \exp(-\delta^2\mu/(2 + \delta)]$ for all $\delta \ge 0$, where $\mu = \E[S^a]$. Choose $\delta$ such that $(1 + \delta)\mu = n/m$. Note that since $a \in L$, $\mu \le \frac{n}{2m}$, so $\delta \ge 1$. Further, this implies that $\delta \mu \ge \frac{n}{2m}$ and $\frac{\delta}{2 + \delta} \le \frac{1}{3}$.  Plugging in both of these bounds, we get an upper bound of $\exp(-\frac{n}{6m})$ on the probability that $a$ wins. Via a union bound, the probability any candidate in $L$ wins is at most $|L| \exp(-\frac{n}{6m})$. \\[-0.5em]

\noindent \textit{Proof of Claim 2.}  Fix two candidates $a, b \in H$. Let $Y_i = X^a_i - X^b_i$; then, we can upper-bound $\Pr[S^a = S^b]$ by $\Pr[\sum_i Y_i = 0]$. Because $Y_i$ converges to be Gaussian-distributed, and since the Gaussian places $0$ mass on any point, we can upper bound $\Pr[\sum_i Y_i = 0]$ by the convergence rate of $\sum_i Y_i$ to the Gaussian, derived via Berry-Esseen: 
\begin{equation} \label{eq:berry-esseen-ub}
    Pr\left[\sum_i Y_i = 0\right] \leq 
 O\left(\nicefrac{1}{\sqrt{\Var[Y_i]}} \cdot  \max_i (\rho_i/\Var[Y_i]) \right)
\end{equation}
 where $\rho_i = \mathbb{E}[|Y_i - \mathbb{E}[Y_i]|^3$. We now bound each of these terms.
 First, since each $Y_i \in [-1, 1]$, $|Y_i - \mathbb{E}{Y_i}| \le 2$, so $\max_i (\rho_i/\Var[Y_i]) \le 2$. 
Next, we expand out $\Var[Y_i]$ to get the following, where the last step uses that $2E[X_i^{a}X_i^{b}] = 0$ since only one of these indicators can be 1 for a given $i$.
\[\Var[Y_i] = \mathbb{E}[Y_i^2] - \mathbb{E}[Y_i]^2 = E[(X_i^{a})^2] + E[(X_i^{b})^2]- 2E[X_i^{a}X_i^{b}] - E[X_i^{a} - X_i^{b}]^2 = p_i^a + p_i^b - (p_i^a - p_i^b)^2,\]

Now, \Cref{lem:first-last}, the probability that \textit{any} candidate $\ranking_i(j)$ ends up in the first position is proportional to $\phi^j$. Therefore, (assuming $m \geq 2$) each $p_i^c$ is upper-bounded by $\frac{\phi}{\sum_{j = 1}^m \phi^j} \le \frac{\phi}{\phi + \phi^2} = \frac{1}{1 + \phi}$. We use this to upper-bound $(p_i^a - p_i^b)^2$ as follows: 
\[(p_i^a - p_i^b)^2 \le \max(p_i^a, p_i^b) (p_i^a + p_i^b) \le \nicefrac{1}{1 + \phi} \cdot (p_i^a + p_i^b).\]
Using this, we conclude that $
\Var[Y_i] \ge p^a_i + p^b_i -  \frac{1}{1 + \phi}(p_i^a + p_i^b) = \frac{\phi}{1 + \phi}(p_i^a + p_i^b).$ It follows that $\sum_i \Var[Y_i] \ge \frac{\phi}{1 + \phi}(\mathbb{E}[S^a] + \mathbb{E}[S^b]) \ge \frac{\phi}{1 + \phi} n/m \ge \frac{\phi}{2} \cdot n/m$. Plugging our bounds into \Cref{eq:berry-esseen-ub},
\[
    Pr\left[\sum_i Y_i=0\right] \leq O\left(\sqrt{m/(n\phi)} \right).
\]
Union bounding over the at most $|H|^2 \leq m^2$ pairs of candidates in $H$, the probability such a tie occurs is at most $O\left(m^2\sqrt{m/(n\phi)}\right)$. Finally, union bounding again over the event that the winner was from $L$ (where $|L| \leq m$), the overall probability of a tie is at most the stated bound.

For $\rho(n)$\textsc{-Group-Stability}, we upper bound the probability that any non-winning candidate has plurality score within $2\rho(n)$ of the maximal score. make the threshold to be in $L$ having $\E[X^c]\le \frac{n}{3m}$. For Claim 1, we use a threshold of $\frac{n}{m} - 2\rho(n) \ge \frac{2n}{3m}$. The same argument goes through, we just plug in $\frac{n}{3m}$ for $\delta\mu$ instead, replacing the $6$ in the bound with a $9$. For Claim $2$, rather than just proving it is unlikely for the scores to be equal, we can do the same analsysis for all differences in the range $[-2\rho(n), 2\rho(n)]$. Union bounding over all $4\rho(n) + 1$ possible score values, this adds an additional $O(\rho(n))$ factor.
\end{proof}

\begin{proposition}[\textsc{Borda}] \label{lem:borda-mallows}
    \textsc{Borda} smoothed-satisfies \textsc{Resolvability} at a rate of $O\left(m^3/\sqrt{n\phi}\right)$, and smoothed-satisfies $\rho(n)-$\textsc{Group-Stability} at a rate of $O\left(\rho(n)m^4/\sqrt{n\phi}\right)$.
\end{proposition}

\begin{proposition}[\textsc{Veto}] \label{lem:veto-mallows}
    The rate \textsc{Veto} smoothed-satisfies \textsc{Resolvability} is lower bounded by $1 - n\phi^{m - 2}$, and, as long as $n \ge m$,  the rate  $1$-\textsc{strategy-proofness} is smoothed-satisfied is lower bounded by $1 - n\phi^{m - 2} - m\phi^{\floor{\frac{n}{m}}}$.
\end{proposition}
\begin{proof}
    Consider a profile where all voters rank two candidates $a$ and $b$ in their top two positions. We upper bound the probability that either $a$ or $b$ is ranked last by any voter; if this never occurs, then $a$ and $b$ will be tied. To this end, recall that by \Cref{lem:first-last}, a voter's first-choice alternative will be ranked last post-noise with probability $\phi^m/\sum_{j = 1}^m \phi^j$, and their second-choice alternative will be ranked last post-noise with probability $\phi^{m - 1}/\sum_{j = 1}^m \phi^j$. Hence, neither of these candidates will ranked last with probability at least
    \begin{align*}
        1 - \frac{\phi^{m - 1} + \phi^m}{\sum_{j = 1}^m \phi^j} = 1 - \frac{\phi^{m -2} + \phi^{m - 1}}{\sum_{j = 0}^1 \phi^j} = 1 - \frac{(\phi^{m -2} + \phi^{m - 1})(1 - \phi)}{1 - \phi^m} = 1 - \frac{\phi^{m -2} -\phi^m}{1 - \phi^m} \ge 1 - \phi^{m - 2}.
    \end{align*}
    Union bounding over all voters, we get that no voter will place $a$ or $b$ last with probability at least $1 - n\phi^{m - 2}$. This lower bounds the rate at which \textsc{Veto} can smoothed-satisfy \textsc{Resolvability}.

    For $1$-\textsc{Group-Stability}, we will have the same construction, except each candidate other than $a$ or $b$ will be ranked last by at least $\floor{\frac{n}{m}}$ voters. It is still the case that $a$ and $b$ will never be ranked last with probability at least $1 - n\phi^{m - 2}$. Note that for any candidate $c \ne a, b$, if a voter ranks them last in the starting profile, they will continue to do so with probability $\frac{1}{\sum_{j = 0}^m \phi^j} = \frac{1 - \phi}{1 - \phi^m} \ge 1 - \phi$. This implies they will continue to be ranked last by at least one of these $\floor{\frac{n}{m}}$ voters with probability at least $1 - \phi^{\floor{\frac{n}{m}}}$. Union bounding over these $m - 2 \le m$ candidates, in the sampled profile, with probability $1 - n\phi^{m - 2} - m\phi^{\floor{\frac{n}{m}}}$, candidates $a$ and $b$ will never be ranked last, while all other candidates will be ranekd last at least once. Hence, $a$ and $b$ are tied as veto winners. Further, as there are at least $m > m - 2$ voters, at least one candidate $c$ must be ranked last at least twice. Consider a voter $i$ that ranked $c$ last, and suppose without loss of generality they prefer $a$ to $b$. In this case, they can move $b$ to the bottom of their ranking. Now $a$ will be the unique candidate never ranked last, and will hence be the unique winner. This is an improvement for the voter, meaning that the resulting profile does not satisfy $1$-\textsc{Strategy-proofness}.
\end{proof}

\begin{proposition}[\textsc{Minimax}] \label{lem:minimax-mallows}
    The rate at which \textsc{Minimax} smoothed-satisfies \textsc{Resolvability} is lower bounded by
    \[\Omega\left( \frac{1}{\sqrt{n\bar{q}(\floor{m/2})}} -  m\exp\left(-\frac{n(1 - 2\bar{q}(\floor{m/2}))^2}{32} \right)\right),\]
    at least for $n$ divisible by $4$, and the rate for $1$-\textsc{Group-Strategyproofness} is lower bounded by \[1 - 3nq(\floor{m/4}) - O\left(m\exp\left(\frac{-(n - 2)(1 - 2\bar{q}(\floor{m/2}))^2}{32} \right)\right).\]
\end{proposition}

Note that using the fact that $\bar{q}(\floor{m/2}) \in O(1/\log m)$, a necessary condition for a high probability bound is $n \in \omega(\bar{q}(\floor{m/2})$. We can plug in our various upper bounds from \Cref{lem:mallows-ubs} on $\bar{q}(\floor{m/2})$. For the bound in \Cref{tab:classification4}, the second of these implies that $n \in \omega(1/(m\phi^{\floor{m/2}}))$ is necessary.

\section{Discussion} \label{sec:discussion}

Because the motivation for the smoothed model is fundamentally practical, we begin by outlining some key practical takeaways from our work. These takeaways help address the question: \textit{how much can a bit of smoothed noise give us, when, and why?}

Although the related work so far has been largely optimistic about smoothed analysis in social choice, our analysis in \Cref{sec:negative-results} illustrates that for many rules and axioms, smoothed noise may be insufficient to circumvent impossibilities. Our sufficient condition for smoothed violation, moreover, suggests that the insufficiency of smoothed noise may be tied fundamentally to axioms' and voting rules' simplicity and interpretability, suggesting that these negative results are difficult to get around without making other compromises.

On the other hand, our results in \Cref{sec:positive-results} show that for certain kinds of axioms\emdash i.e., those whose counterexamples must lie near hyperplanes or other small-measure structures\emdash a small amount of noise \textit{is} enough to circumvent impossibilities. However, our results in \Cref{sec:beyond-minprob} paint a more nuanced picture: if we believe that in practice, any perturbations are truly unlikely to cause drastic opinion changes, then our choice of voting rule is far more important than past work suggests. In particular, while past $\minprob$-parameterized upper bounds yield similar convergence rates across voting rules (for any given axiom), our results in the Mallows model\emdash for which existing bounds do not usefully apply\emdash show that actually, certain voting rules converge to smoothed-satisfaction far faster than others. The voting rules that converge faster are, perhaps unsurprisingly, those whose outcomes are more sensitive to local swaps.

With these takeaways in hand, we now explore some extensions and future work.\\\vspace{-.5em}

\noindent\textbf{Extension and future work: smoothed analysis of Arrow's Theorem} 
Our results from \Cref{sec:negative-results} and \ref{sec:positive-results}, which address the smoothed-satisfaction of individual axioms, also have implications for \textit{multi-axiom impossibilities} of the form, ``there exists no voting rule that simultaneously satisfies all axioms in $A \in \mathcal{A}$'' where $\mathcal{A}$ is some collection of axioms. We can now ask the same question replacing ``satisfies'' with ``smoothed-satisfies.'' %
Past work in the semi-random model has studied several multi-axiom impossibilities, and our results from \Cref{sec:results1} immediately recover several of these existing results.\footnote{Our results imply the smoothed-resolution Gibbard-Satterthwaite \cite{gibbard1973manipulation} (\Cref{cor:GS}), the ANR impossibility theorem (e.g., see \cite{xia2020smoothed}) (\Cref{cor:ANR}), and the impossibility of simultaneously satisfying \textsc{Condorcet} and \textsc{Participation} as identified by Moulin \cite{moulin1988condorcet} (\Cref{cor:condpart}). Our results also imply that the existence of Condorcet cycles (i.e., \textit{Condorcet's Paradox}) is \textit{not} smoothed-resolved\emdash that is, existing Condorceet cycles remain with high probability in the presence of smoothed noise (\Cref{prop:condcycles}).} However, perhaps the most famous multi-axiom impossibility, \textit{Arrow's Theorem} \cite{arrow1951individual}, has yet to be analyzed under semi-random noise. This impossibility states that there is no voting rule that (when $m \geq 3$) can simultaneously satisfy \textsc{IIA}, \textsc{Unanimity}, and \textsc{Non-Dictatorship.}\footnote{\textsc{Unanimity} requires that if an alternative is ranked first by all voters, it is the winner. \textsc{Non-Dictatorship} means that there exists no voter such that the outcome of the voting rule is always their first choice.} 

In light of our findings in \Cref{sec:negative-results} that \textsc{IIA} tends to be smoothed-violated, one might guess that Arrow's Impossibility is smoothed-impossible; i.e., it still holds under smoothed noise. However, due to how Arrow defines the axiom \textsc{Non-dictatorship} \cite{arrow1951individual}, it is not hard to show that Arrow's Theorem is in fact \textit{smoothed-resolved}. Indeed, a voting rule is said to be a \textit{dictatorship} if there exists some voter such that the outcome of the voting rule is always that voter's first choice. Now define an \textit{almost-dictatorship} voting rule, that chooses voter 1's favorite alternative on every profile except one arbitrary profile. The existence of this single exceptional profile means that our rule satisfies \textsc{Non-dictatorship}. However, since the rule only differs from a dictatorship on this one profile, it can easily be checked that it smoothed-satisfies \textsc{IIA} and \textsc{Unanimity}.

This positive result for Arrow's theorem is not very conceptually satisfying, as it does not get at the heart of what makes IIA, dictatorship, and unanimity inconsistent. In some sense, this is because satisfying non-dictatorship, as it is defined, is too easy. To say something more meaningful, we propose a strengthening of \textsc{Non-Dictatorship}, called \textsc{Local Non-Dictatorship}, that may be more interesting in the smoothed context: 
Fix a profile $\profile$ and define voter $i$'s neighborhood around $\profile$, $N_i(\profile) \subseteq \profiles$, to be the set of all profiles reachable by switching $i$'s ranking for another ranking. Then, we say rule $R$ satisfies \textsc{Local Non-Dictatorship} on profile $\profile$, if for all voters $i$, there is a $\profile' \in N_i(\profile)$ such that $i$'s first choice in $\profile'$ doesn't win. In a sense, then, satisfying this axiom disallows a voter from being a dictator in any \textit{local} area of profile space. 

Because satisfaction of \textsc{Local Non-Dictatorship} implies the satisfaction of \textsc{Non-Dictatorship}, Arrow's impossibility also implies inconsistency between \textsc{IIA, Unanimity}, and \textsc{Local Non-Dictatorship}. We refer to this new, stronger impossibility as \textit{
Strengthed Arrow's Theorem.}
As we hoped, Strengthened Arrow's Theorem is not susceptible to the simple fix discussed above: our \textit{almost-dictatorship} voting rule fails to smoothed-satisfy \textsc{Local Non-Dictatorship}, since voter 1 is the dictator across almost every neighborhood.
Then, the question remains open: \textit{can this impossibility be smoothed-resolved?}\\\vspace{-.5em}

\noindent\textbf{Future work: more general models limiting drastic opinion change}
Our analyses in \Cref{sec:beyond-minprob} are restricted to the Mallows model, which is perhaps the poster child for a much broader, practically-motivated assumption: \textit{noise-inducing shocks should not, in practice, induce drastic opinion change.}
Already, our results reveal that this is a practically meaningful restriction, uncovering previously hidden heterogeneity among voting rules' rates of convergence, due to their varying sensitivity to local swaps. 

Given the practicality of the above assumption\emdash and the different patterns of convergence that emerge when we apply it\emdash we see investigating the broader space of noise models satisfying this assumptoin as an important frontier. There are many possible such noise models, and many will have $\minprob$ equal to or very near zero, warranting the pursuit of new bounds. Such further analyses may enable more nuanced distinctions between axioms and voting rules, and stronger conclusions over more flexible noise models. Going even further, one exciting potential opportunity in this direction is to potentially trade this restriction for a relaxation of the \textit{independence} assumption (that noise is applied independently across rankings). This strong assumption---relied upon centrally in both our analyses and those in the related work---is difficult to relax when reasoning about almost arbitrary noise distributions. However, there may be hope of relaxing it when considering noise distributions restricted to more closely reflect practice.

\newpage
\bibliographystyle{plainnat}
\bibliography{abb,bibliography}

\newpage
\appendix
\section{Supplemental Materials from \Cref{sec:introduction}}

\subsection{Comparison to Spielman and Teng's smoothed model} \label{app:st_model} The smoothed analysis framework was originally proposed by Spielman and Teng to provide theoretical justification for the Simplex algorithm's fast runtime in real-world instances, despite its exponential worst-case complexity~\cite{spielman2004smoothed}. In their analysis, they go beyond the worst-case by adding Gaussian noise independently to each entry of the real-valued constraint matrix that is the input to Simplex algorithm. Then, they bound the expected run time of Simplex in the worst-case over inputs, where these guarantees are parameterized by the variance of the Gaussian noise added. 
Stated as a more general framework, the idea of smoothed analysis is to fix an arbitrary instance, add noise from a parameterized distribution, and then measure the quality of the expected outcome on the worst input (or, whether a property is satisfied with high probability---an alternative formulation proposed by Spielman and Teng that is closer to ours).  

This is precisely how our model works, but rather than its input being a constraint matrix of real numbers, it is a \textit{base profile} of complete rankings over $m$ alternatives. This base profile is perturbed by applying generically-structured noise independently across each of its rankings (we refer to this as the \textit{independence} assumption). When evaluating the probability of a criterion being satisfied post-perturbation, we assume that the base profile is chosen adversarially, i.e., to minimize this probability. The noise distribution we apply to each ranking is parameterized by a value $\phi \in [0,1]$ to measure the quantity of noise added, analogous to Spielman and Teng's variance parameter. Our main departure from the original smoothed model is, where Spielman and Teng specify Gaussian noise, we do not assume a specific noise distribution, instead allowing \textit{any} $\phi$-parameterized distribution that is \textit{neutral} over alternatives.\footnote{We do not commit to a noise distribution because there is no single well-established distribution that is obvious to apply (unlike in the real-valued setting, where the Gaussian is standard).} \textit{Neutrality} means that for a given $\phi$, if we permute a ranking and add noise, this is equivalent to adding noise and then permuting the output. In that sense, our noise model over profiles can be specified by a single distribution over one ranking, permuted to be applied to ranking permutations. On this distribution over rankings, we also sometimes assume \textit{positivity}---that when $\phi > 0$, this distribution assigns positive probability to all rankings. Under these assumptions,
Our class of noise models generalizes the popular Mallows noise model (e.g.,~\cite{lu2011learning}) as well as one-dimensional parameterizations of the Plackett-Luce model (e.g., see~\cite{cheng2010label}).\footnote{Traditionally, the Plackett-Luce model takes a single $m$ real-valued parameters, one per ranking position. Our model generalizes a variation where each parameter is expressed as a function of $\phi$.}

\subsection{Comparison of techniques with Xia 2020 \cite{xia2020smoothed}} \label{app:techniques}
At a high level, Xia's work and our work in the general smoothed model (\Cref{sec:results1}) take a similar technical approach: both show that their noise models become well-behaved as the number of voters $n$ grows large, and then use these convergence results to upper and lower bound how much probability mass is placed on ``bad'' profiles. 
In both models, the key assumption enabling this type of analysis is the independence of noise across rankings. As a result of this assumption, as $n$ grows large, the resulting distribution over profiles will converge to distributions we understand, analogous to how sums of i.i.d.\@ random variables converge to Gaussians via the Central Limit Theorem, or concentrate around their mean via Hoeffding's inequality. We prove the convergence of our model in \Cref{lem:starting-concetration,lem:normality}, corresponding to Xia's Lemma 1 of~\cite{xia2020smoothed}.\footnote{Notice that when comparing formal statements, both models consider the space of histograms (vectors representing the fraction of the profile composed of each ranking) rather than profiles directly, and, while we tend to analyze the set of histograms corresponding to ``bad'' profiles (e.g., those in which an axiom is not satisfied) directly, Xia considers sets which are solutions to a system of linear equations and inequalities. However, in practice, these approaches end up being quite similar. 
} Our convergence rates across the respective lemmas match asymptotically except when roughly, the set of ``bad'' profiles for a criterion is extremely small; this case tends to only come into effect for very specific kinds of criterion, none of which we formally analyze in this paper. 
In the limited cases where such differences do exist, they stem from our proof relying on a multi-dimensional version of the well-known Berry-Esseen theorem, while Xia's relies on faster convergence of Poisson Multinomial variables.

\newpage
\section{Supplementary Materials from \Cref{sec:model}}\label{app:model}
\subsection{Definitions of Voting Rules} \label{app:rules}
We define the voting rules we study as functions of profile histograms $\proportions$ (rather than a profile $\profile$). Further, we define them in the same form: first, we express how they assign candidates' scores, and then express via an arg$\max_c$ that the winner or set of winners constitutes the candidate(s) with the highest (or in one case, lowest) score.

\paragraph{Positional Scoring Rules (PSRs).} For fixed $m$, a positional scoring rule is characterized by a vector of weights of length $m$, $\boldsymbol{\alpha} = (\alpha_c | c \in [m])$, where $\alpha_1 \geq \alpha_2 \geq \dots \geq \alpha_m$. Without loss of generality, we let these weights be translated and scaled such that $\alpha_1 = 1$ and $\alpha_m = 0$. The winner(s) by a PSR $R$, characterized by $\boldsymbol{\alpha}$,
 \begin{align*}
      R(\proportions) &= \text{arg}\max_c \sum_{j \in [m]} \alpha_j \sum_{\substack{\ranking \in \rankings \\ \ranking(j) = c}} \proportion_\ranking
 \end{align*}

\paragraph{\textsc{Minimax}.} The \textsc{Minimax} winner is the candidate whose greatest pairwise defeat is the smallest:
 \begin{equation*}
     \textsc{Minimax}(\proportions) =  \text{arg}\min_c \max_{c' \neq c} \sum_{\substack{\ranking \in \rankings \\ c'\succ_\ranking c}} \proportion_\ranking
 \end{equation*}

\paragraph{\textsc{Kemeny-Young}.} We define a candidate $c$'s Kemeny-Young score to be, at a high level, the level of agreement with voters' rankings of the most-agreeing ranking that ranks $c$ first. Then, the set of winners contains the candidate(s) with the highest Kemeny-Young score.
 \begin{equation*}
     \textsc{Kemeny-Young}(\proportions) = \text{arg}\max_c \max_{\ranking : \ranking(1) = c}\sum_{\substack{c',c'' \in C \\ c' \succ_\ranking c''}} \sum_{\substack{\ranking' \in \rankings \\ c' \succ_{\ranking'} c''}} \proportion_{\ranking'}
 \end{equation*}

\paragraph{\textsc{Copeland}.} 
 \[\textsc{Copeland}(\proportions) = \text{arg}\max_c \sum_{c' \neq c} \mathbb{I}[c \succ_{\proportions} c'] +  \nicefrac{1}{2} \cdot \mathbb{I}[c \sim_{\proportions} c'] \]

\subsection{Definitions of Axioms} \label{app:axioms}
We use the following notation. For a ranking $\ranking \in \rankings$, we let $\ranking(j)$ for an index $j \in [m]$ be the candidate in the $j$'th position of $\ranking$. For a ranking $\ranking \in \rankings$ and distinct candidate $c, c' \in \cands$, we use $c \succ_\ranking c'$ to denote that $c$ is ranked higher than $c'$ in $\ranking$. $c \succ_\profile c' \text{ when } |\set{i \in [n] \suchthat c \succ_{\ranking_i} c'}| > |\set{i \in [n] \suchthat c' \succ_{\ranking_i} c}|$.\\

\noindent \textsc{Resolvability}.
A voting rule $R$ satisfies \textsc{Resolvability} on $\profile$ iff $|R(\profile)| = 1$ (i.e., there are no ties).\\

\noindent \textsc{Condorcet}. A \textit{Condorcet winner} is a candidate that would win in a pairwise election against every other candidate. That is, $c$ is a Condorcet winner in $\profile$ if  $c\succ_{\profile} c'$ for all $c' \neq c$.
A voting rule $R$ satisfies Condorcet Consistency on a given profile $\profile$ if one of two conditions hold: (1) there is no Condorcet winner in $\profile$, or (2) there is a Condorcet winner $c$, and $R(\profile) = \set{c}$.\\

\noindent\textsc{Majority}.
A \textit{majority winner} is a candidate that is ranked first by a majority of agents. That is, $c$ is a majority winner in $\profile$ if
\[|\set{i \in [n] \suchthat \ranking_i(c) = 1}| > n/2\]
A voting rule $R$ satisfies \textsc{Majority} on a profile $\profile$ if it satisfies one of two conditions: (1) there is no majority winner in $\profile$, or (2) if there is a majority winner $a$, then $R(\profile) = \set{a}$.\\

\noindent\textsc{Consistency}. A voting rule $R$ satisfies \textsc{Consistency} on a profile $\profile$ if the following holds: for all partitions of $\profile$ into sub-profiles, $(\profile^1, \ldots, \profile^t)$, if $R(\profile^j)$ for all $j \in [t]$ is the same set of winners $W$, then $R(\profile) = W$. \\

\noindent \textsc{Independence of Irrelevant Alternatives (IIA)}. A voting rule $R$ satisfies \textsc{Independence of irrelevent alternatives (IIA)} on profile $\profile$ if the following holds. Suppose $R(\profile) = a$. Then, for all candidates $b \ne a$, if $\profile'$ is such that $a \succ_{\ranking_i} b$ if and only if $a \succ_{\ranking_i'} b$ for all voters $i$, then $R(\profile') \ne b$.\\

\noindent The axioms we study are defined formally as follows:

\begin{definition}[\textbf{$\boldsymbol{\rho(n)}$\textsc{-group-strategyproofness}}] A voting rule $R$ satisfies $\rho(n)$\textsc{-group-strategy- proofness} on a profile $\profile$ if there exists no group of agents of size at most $\rho(n)$ such that if they change their votes, resulting in some profile $\profile'$ with outcome $R(\profile')$, they are all at least as well off and at least one is strictly better off. An agent is at least as well (resp. strictly better) off if their favorite candidate in the set $R(\profile')$ is weakly (resp. strictly) preferred to their favorite candidate in $R(\profile)$.
\end{definition}

\begin{definition}[\textbf{$\boldsymbol{\rho(n)}$\textsc{-group-monotonicity}}] A voting rule $R$ satisfies $\rho(n)$\textsc{-group-monotonicity} on a profile $\profile$ if there exists no candidate $a$ and no group of agents of size at most $\rho(n)$ such that if they change their votes \textit{without decreasing the position of $a$ in any of their rankings}, producing a new profile $\profile'$, then it cannot be that $a \in R(\profile)$ and $a \notin R(\profile')$.
\end{definition}

\begin{definition}[\textbf{$\boldsymbol{\rho(n)}$\textsc{-group-participation}}] A voting rule $R$ satisfies $\rho(n)$\textsc{-group-participation} on a profile $\profile$ if there exists no group of agents of size at most $\rho(n)$ such that if they collectively leave the election, producing a new profile $\profile'$, they are all at least as well off and at least one is strictly better off with the outcome $R(\profile')$ than $R(\profile)$.
\end{definition}

\newpage
\section{Supplementary material from \Cref{sec:results1}}
\subsection{Proof of \Cref{prop:CSRS_axioms}}\label{app:CSRS_axioms}
We prove this result for each axiom separately. For each axiom, we define a sufficient condition for a counterexample $\profile$ to be ``strict'', i.e., all profiles $\profile'$ with histograms nearby to $\proportions^\profile$ will also be counterexamples. We then show that the existence of a strict counterexample implies smoothed-violation via \Cref{thm:violation-sufficient}. We later give (or point to existing) strict counterexamples for all relevant pairs of rules and axioms. In the following arguments, we say a profile $\profile$ is \emph{robust} with respect to a voting rule $R$ if there is an $\e > 0$ such that all profiles $\profile'$ with $\| \proportions^{\profile} - \proportions^{\profile'} \|_1 < \e$, $R(\profile') = R(\profile)$. Notice that for hyperplane rules, all profiles that do not fall on hyperplanes are robust.

\paragraph{\textsc{Condorcet}:} We say a counterexample $\profile$ is a strict counterexample to $R$ satisfying \textsc{Condorcet} if $\profile$ is robust with respect to $R$, $\profile$ has a strict Condorcet winner $a$ (i.e., a candidate that beats every other candidate on \emph{strictly more} than half of the voters), and $R(\profile) \ne a$. If such a strict counterexample $\profile$ exists, if it is robust with value $\e_1$ and $a$ wins with at least a $1/2 + \e_2$ fraction against each candidate, all profiles whose histogram falls within $\e = \min(\e_1, \e_2)$ of $\proportions^\profile$ have the same Condorcet winner and same output under $R$, and thus are also counterexamples. Hence, \Cref{thm:violation-sufficient} implies smoothed-violation.

\paragraph{\textsc{Majority}:}We say a counterexample $\profile$ is a strict counterexample to $R$ satisfying \textsc{Majority} if $\profile$ is robust and $\profile$ has a strict Majority winner $a$ (i.e., a candidate that is ranked first by \emph{strictly more} than half of the voters), and $R(\profile) \ne a$. If such a strict counterexample $\profile$ exists, if it is robust with value $\e_1$ and $a$ is ranked first by at least a $1/2 + \e_2$ fraction of the voters, all profiles whose histogram falls within $\e = \min(\e_1, \e_2)$ of $\proportions^\profile$ have the same Condorcet winner and same output under $R$, and thus are also counterexamples. Hence, \Cref{thm:violation-sufficient} implies smoothed-violation.

\paragraph{\textsc{Consistency}:} We say a counterexample $\profile$ is a strict counterexample to $R$ satisfying \textsc{Consistency} if $\profile = \profile^1 \cup \cdots \cup \profile^t$ such that $R(\profile^1) = \cdots = R(\profile^t) \ne R(\profile)$ and all of $\profile^1, \ldots, \profile^t$ and $\profile$ are robust with respect to $R$. Suppose such a strict counterexample $\profile = \profile^1 \cup \cdots \cup \profile^t$ exists. Let $\e^{min}$ be an amount by which all the relevant profiles are robust. Suppose each profile $\profile^j$ has $n_j$ voters and let $n = n_1 + \cdots + n_t$ be the number of voters in $\profile$. Let $p = \min_j n_j / n$. Notice that any $\profile'$ on $zn$ voters within $\e = p\e^{min}$ of $\profile$ can be decomposed into $\profile^{1'}, \ldots, \profile^{t'}$ such that each $\profile^{j'}$ has $zn_j$ voters and $\proportions^{\profile^{j'}}$ is at most $\e^{min}$ away from $\proportions^{\profile^j}$. Hence, $R(\profile^{1'}) = \cdots = R(\profile^{t'}) \ne R(\profile')$, so $\profile'$ is a counterexample to \textsc{Consistency}. Hence, \Cref{thm:violation-sufficient} implies smoothed-violation.

\paragraph{\textsc{IIA}:} We say a counterexample $\profile^1$ is a strict counterexample to $R$ satisfying \textsc{IIA} if there is another profile  $\profile^2$ such that both $\profile^1$ and $\profile^2$ are robust with respect to $R$, and there are candidates $a, b$ such that the relative ranking of $a$ and $b$ are the same under $\ranking^1_i$ and $\ranking^2_i$ for all voters $i$, yet $R(\profile^1) = a$ and $R(\profile^2) = b$. Suppose such a strict counterexample $\profile^1$ and $\profile^2$ that are both robust by at least $\e$. Consider any profile $\profile^{1'}$ where $\proportions^{\profile^{1'}}$ is within $\e$ of $\proportions^{\profile^1}$. Notice that there must be a profile $\profile^{2'}$ such that $\proportions^{\profile^{2'}}$ is within $\e$ of $\proportions^{\profile^{2}}$ that matches $\profile^{1'}$ in terms of all voter's relative rankings of $a$ and $b$, and by robustness $R(\profile^{1'}) = a$ while $R(\profile^{2'}) = b$. Hence, \Cref{thm:violation-sufficient} implies smoothed-violation.

\paragraph{Counterexamples:}
The following table points to strict counterexamples that can be used to show the claims above. Most can be found on Wikipedia pages. The missing ones are presented afterwards.

\begin{table}[ht]
    \centering
    \begin{tabular}{r||ccccc}
        \multicolumn{1}{l}{} & \multicolumn{5}{c}{\textbf{Axioms}}\\
         \textbf{Voting Rules} &   \textsc{Condorcet}  &  \textsc{Majority} &   \textsc{Consistency} & \textsc{IIA} &\\
         \hline \hline
        \textsc{Plurality} &  \cellcolor{gray!10} \textbf{[a]} &  \cellcolor{gray!10} \textcolor{gray}{satisfied} &  \cellcolor{gray!10} \textcolor{gray}{satisfied} &  \cellcolor{gray!10} \textbf{[c]} &  \\
        (non-\textsc{Plurality}) PSRs &  \cellcolor{gray!10} \textbf{\Cref{ex:PSRs_cond}} & \cellcolor{gray!10} \textbf{\Cref{ex:PSRs_cond}} & \cellcolor{gray!10} \textcolor{gray}{satisfied} &  \cellcolor{gray!10} \textbf{\Cref{ex:PSRs_IIA}}  & \\
        \textsc{Minimax} & \cellcolor{gray!10}   \textcolor{gray}{satisfied} & \cellcolor{gray!10}  \textcolor{gray}{satisfied} & \cellcolor{gray!10}  \textbf{[b]} &  \cellcolor{gray!10} \textbf{[c]} &     \\
        \textsc{Kemeny-Young}  &   \cellcolor{gray!10} \textcolor{gray}{satisfied} &  \cellcolor{gray!10} \textcolor{gray}{satisfied} &   \cellcolor{gray!10}  \textbf{[b]} &  \cellcolor{gray!10} \textbf{[c]} &  \\
        \textsc{Copeland}  &   \cellcolor{gray!10} \textcolor{gray}{satisfied} &  \cellcolor{gray!10} \textcolor{gray}{satisfied} &   \cellcolor{gray!10}  \textbf{[b]} &  \cellcolor{gray!10} \textbf{[c]} &  \\
    \end{tabular}
    \caption{Externally-referenced examples come from the following Wikipedia pages [a]: \textit{Condorcet Criterion}, [b]: \textit{Consistency Criterion}, [c]: \textit{Independence of Irrelevant Alternatives}.}
    \label{tab:app_classification}
\end{table}

\begin{example} \label{ex:PSRs_cond}
Let $R$ be the positional scoring rule represented by the weights vector $(1,\alpha, \dots,0)$ without loss of generality. We assume $\alpha>0$ is separated from 0 (otherwise, $R$ is just Plurality). Let
$$\profile: 
\begin{tabular}{l|l}
     $n \left(\frac{1}{2} - \frac{\alpha}{4(2-\alpha)}\right)$ voters & $\boldsymbol{a_1} \succ a_3 \succ \dots \succ a_m \succ \boldsymbol{a_2}$ \\
     $n \left(\frac{1}{2} + \frac{\alpha}{4(2-\alpha)}\right)$ voters  & $\boldsymbol{a_2} \succ \boldsymbol{a_1} \succ a_3 \succ \dots \succ a_m$ \\
\end{tabular}. 
$$
\end{example}

\begin{example}\label{ex:PSRs_IIA}
Let $R$ be the positional scoring rule represented by the weights vector $(1,\alpha, \dots,0)$ without loss of generality. We assume $\alpha>0$ is separated from 0 (otherwise, $R$ is just Plurality). 
$$\profile^1: 
\begin{tabular}{l|l}
    $n/2$ voters  & $\boldsymbol{a_1} \succ a_3 \succ \dots \succ a_m \succ \boldsymbol{a_2}$ \\
     $n/4$ voters & $\boldsymbol{a_2} \succ a_3 \succ \dots \succ a_m \succ \boldsymbol{a_1}$ \\
     $n/4$ voters  & $\boldsymbol{a_2} \succ \boldsymbol{a_1} \succ a_3 \succ \dots \succ a_m$ \\
\end{tabular} \hspace{0.75cm} \profile^2:
\begin{tabular}{l|l}
    $n/2$ voters  & $\boldsymbol{a_1} \succ \boldsymbol{a_2} \succ a_3 \succ \dots \succ a_m $ \\
     $n/4$ voters & $\boldsymbol{a_2} \succ a_3 \succ \dots \succ a_m \succ \boldsymbol{a_1}$ \\
     $n/4$ voters  & $\boldsymbol{a_2} \succ \boldsymbol{a_1} \succ a_3 \succ \dots \succ a_m$ \\
\end{tabular}.
$$
\end{example}

This concludes the proof. \qed

\subsection{Proof of \Cref{lem:starting-concetration}}\label{app:starting-concentration}
Fix a noise model $\noisemodel$ and $\e > 0$. By continuity of $\noisemodel$, there exists a $\phi >0$ such that for all $\phi' \in [0, \phi]$, $\Pr[\noisemodel_{\phi'}(\ranking) = \ranking] > 1 - \e/2$. Choose this to be our $\phi$.

Fix such a $\phi'$ and a profile $\profile \in \profiles_n$. Notice that after applying $\noisemodel$, each ranking $\ranking_i$ will stay the same with probability at least $1 - \e / 2$. As this is independent accross voters, a straightforward application of Hoeffding's inequality tells us that at least a $1 - \e$ fraction of rankings will not change with probability at least $1 - \exp(\e^2 n / 2)$, as needed.
\qed

\subsection{Proof of \Cref{prop:resolvability}}\label{app:resolvability}
For a decisive hyperplane rule, note that all profiles that fail  \textsc{Resolvablity} must fall on one of the $\ell$ hyperplanes (where by definition $\ell$ is finite). Further, note that these hyperplanes are convex and measure-zero. Hence, we can immediately apply \Cref{thm:brittle_convex} using the hyperplanes as convex sets.

For non-decisive hyperplane rules, there must exist a profile $\profile$ not lying on any hyperplane for which  \textsc{Resolvablity} fails. Further, $\proportions^\profile$ is at least some $L^1$ distance $\e > 0$ from all hyperplanes. All such profiles with histograms in this ball have the same outcome as $\profile$ and hence do not satisfy \textsc{Resolvablity}. We can then directly apply \Cref{thm:violation-sufficient} using this profile $\profile$ and $r = \e$.

\subsection{Proof of \Cref{lem:normality}}\label{app:normality}
To prove this, we first prove the following technical lemma. One takeaway from this lemma it that $\propdistprof$'s covariance matrix is quite easy to work with: its inverse not only exists but has a simple closed-form, and its eigenvalues are lower-bounded by a constant. 
\begin{lemma}\label{lem:covariance}
    For all noise models $\noisemodel$, parameters $\phi \in (0, 1]$, and rankings $\profile \in \profiles_n$, the covariance matrix $\Cov[\propdistprof]$ is invertible and has all positive real eigenvalues lower bounded by $\minprob(S_\phi)/(m! n)$.
\end{lemma}
\begin{proof}We first express the expectation and variance of $\propdistprof$ in terms of the analogous values for the $\propdistrankingi$. The relationships between these quantities are shown below, derived by applying simple properties of the expectation and variance in conjunction with the fact that $\propdistprof = \nicefrac{1}{n} \sum_{i = 1}^n \propdistrankingi$:
\begin{equation*}
    \E[\propdistprof] = \nicefrac{1}{n} \sum_{i = 1}^n \E[\propdistrankingi] \label{eqn:exp_dist_agents}, \qquad \Cov[\propdistprof] = \nicefrac{1}{n^2} \sum_{i = 1}^n \Cov[\propdistrankingi]
\end{equation*}
We now use these relationships to find closed forms for each of these objects. Note that $\E[\propdistrankingi]$ is an $|\rankingsmissing|$-length vector whose $\ranking$-th component is simply $\Pr[\rankingdistrankingi = \ranking]$. $\Cov[\propdistrankingi]$ is a $|\rankingsmissing| \times |\rankingsmissing|$ matrix whose entries each correspond to a pair of rankings $\ranking,\ranking'$, such that the $\ranking,\ranking'$-th entry is equal to the covariance between the random variables $\propdistrankingi_\ranking$ and $ \propdistrankingi_{\ranking'}$.

The covariance matrix $\Cov[\propdistrankingi]$ is a $|\rankingsmissing| \times |\rankingsmissing|$ matrix whose entries each correspond to a pair of rankings $\ranking,\ranking'$, such that the $\ranking,\ranking'$-th entry is equal to the covariance between the random variables $\propdistrankingi_\ranking$ and $ \propdistrankingi_{\ranking'}$. Given that $\propdistrankingi$ can take on the values of only basis vectors, the values of these random variables are either 0 or 1. To characterize these entries, we will use the fact that for general $X$ and $Y$, $\Cov(X, Y) = \E[XY] - \E[X]\E[Y]$.

For distinct rankings $\ranking \ne \ranking'$, at most one of $\propdistrankingi_\ranking$ and $ \propdistrankingi_{\ranking'}$ can be nonzero, so the expectation their product must be $0$. Then, 
$$\Cov\big(\propdistrankingi_\ranking,\, \propdistrankingi_{\ranking'}\big) = 0 -\Pr[\rankingdistrankingi = \ranking] \cdot \Pr[\rankingdistrankingi = \ranking'].$$ For diagonal entries where $\ranking = \ranking'$, since the values of our random variables are always $0$ or $1$, we have
$$\Cov\big(\propdistrankingi_\ranking,\, \propdistrankingi_{\ranking}\big) = \Pr[\rankingdistrankingi = \ranking] - \Pr[\rankingdistrankingi = \ranking]^2.$$

With these in hand, we now prove the lemma statement. Fix $\noisemodel, \phi,$ and $\profile$. Recall that $\Cov[\propdistprof] = \nicefrac{1}{n^2} \sum_{i = 1}^n \Cov[\propdistrankingi]$. Hence, we first consider $\Cov[\propdistranking]$ for individual rankings $\ranking$.

Fix an arbitrary ranking $\ranking$. First, we will prove that all eigenvalues of $\Cov[\propdistranking]$ are positive and the minimum is at least $\minprob(\noisemodel_\phi)/m!$. 

To simplify notation in the subsequent computations, for the $j$'th ranking $\ranking'$, let $q_j = \Pr[\rankingdistranking = \ranking']$. Recall that in $\Cov[\propdistranking]$, the $(j, k)$-th entry when $j = k$ (a diagonal entry) has value $q_j(1 - q_j)$ and for $j \ne k$, the entry has value $-q_j \cdot q_k$. We can then write the covariance matrix as
$$\Cov[\propdistranking] = \begin{pmatrix}
        q_1(1 - q_1) & -q_1 \cdot q_2 & \cdots & -q_1 \cdot q_{m! - 1}\\
        -q_2 \cdot q_1 & q_2(1 - q_2) & \cdots & \vdots\\
        \vdots & \ddots & \ddots & \vdots \\
        -q_{m! - 1} \cdot q_1 & \cdots & \cdots & q_{m! - 1}(1 - q_{m! - 1})
    \end{pmatrix}.$$
Note that $q_{m!}$ is the probability of the ``missing'' ranking, and that $\sum_{j = 1}^{m!} q_j = 1$. We first demonstrate an inverse of $\Cov[\propdistranking]$. Consider the matrix:
$$M^{inv} = \begin{pmatrix}
       \frac{1}{q_1} + \frac{1}{q_{m!}} &  \frac{1}{q_{m!}} & \cdots & \frac{1}{q_{m!}}\\
        \frac{1}{q_{m!}} & \frac{1}{q_2} + \frac{1}{q_{m!}} & \cdots & \vdots\\
        \vdots & \ddots & \ddots & \vdots \\
          \frac{1}{q_{m!}} & \cdots & \cdots & \frac{1}{q_{m! - 1}}  + \frac{1}{q_{m!}}
    \end{pmatrix}.$$
More formally, the $j$th diagonal entry is $1/q_j - 1/q_{m!}$ and all off-diagonal entries are simply $1/q_{m!}$. 

We now show that $\Cov[\propdistranking] \cdot M^{inv} = I_{m! - 1}$ where $I_{m! - 1}$ is the identity matrix, that is, $M^{inv}$ is in fact the inverse of $\Cov[\propdistranking]$. To that end, let us consider the $i$'th diagonal entry of the product. It is precisely
\begin{align*}
    (\Cov[\propdistranking] \cdot M^{inv})_{jj}
    &=\sum_{k = 1, k \ne j}^{m! - 1} -\frac{q_j q_k}{q_{m!}} + q_j(1 - q_j)\left(\frac{1}{q_j} + \frac{1}{q_{m!}}\right)\\
    &= \sum_{k = 1, k \ne j}^{m! - 1} -\frac{q_j q_k}{q_{m!}} + (1 - q_j) +  \frac{q_j}{q_{m!}}(1 - q_j)\\
    &= \sum_{k = 1, k \ne j}^{m! - 1} -\frac{q_j q_k}{q_{m!}} + 1 - q_j +  \frac{q_j}{q_{m!}} - \frac{q_j \cdot q_j}{q_{m!}}\\
    &= \sum_{k = 1}^{m! - 1} -\frac{q_j q_k}{q_{m!}} + 1 - q_j +  \frac{q_j}{q_{m!}}\\
    &=\frac{-q_j (1 - q_{m!})}{q_{m!}} + 1 - q_j +  \frac{q_j}{q_{m!}}\\
    &=\frac{-q_j + q_j \cdot q_{m!}}{q_{m!}} + 1 - \frac{q_j \cdot q_{m!}}{q_{m!}} +  \frac{q_j}{q_{m!}}\\
    &= 1.
\end{align*}
For a non-diagonal entry $j, k$ with $j \ne k$, we have
\begin{align*}
     (\Cov[\propdistranking] \cdot M^{inv})_{jk}
     &= \sum_{\ell = 1, \ell \ne j, k}^{m!}-\frac{q_j q_\ell}{q_{m!}}  + \frac{q_j(1 - q_j)}{q_{m!}} - q_jq_k \cdot \left(\frac{1}{q_k} + \frac{1}{q_{m!}}\right)\\
     &= \sum_{\ell = 1, \ell \ne j, k}^{m!}-\frac{q_j q_\ell}{q_{m!}}  + \frac{q_j}{q_{m!}} - \frac{q_j q_j}{q_{m!}} - q_j - \frac{q_jq_k}{q_{m!}}\\
     &= \sum_{\ell = 1}^{m!}-\frac{q_j q_\ell}{q_{m!}}  + \frac{q_j}{q_{m!}}  - q_j\\
     &= \frac{q_j(1 - q_{m!})}{q_{m!}} + \frac{q_j}{q_{m!}}  - \frac{q_j q_{m!}}{q_{m!}}\\
     &= 0.
\end{align*}
We now consider the eigenvalues of $\Cov[\propdistranking]$. Since it is a covariance matrix, it is symmetric, and therefore positive semi-definite. Since we now know it is invertible, it is in fact positive definite. This implies all of its eigenvalues exist and are positive. Further, since the eigenvalues of $M^{inv} = \Cov[\propdistranking]^{-1}$ are the reciprocals of the eigenvalues of $\Cov[\propdistranking]$, we can lower bound the eigenvalues of $\Cov[\propdistranking]$ by upper bounding the the eigenvalues of $M^{inv}$. 

To upper bound the maximum eigenvalue of $M^{inv}$, we can upperbound the maximum absolute row sum. Note that the sum of row $j$ is $$\frac{1}{q_j} + \frac{m! - 1}{q_{m!}} \le \frac{m!}{\minprob(S_\phi)}.$$
This lower bounds the minimum eigenvalue of $\Cov[\propdistranking]$ by $\frac{\minprob(S_\phi)}{m!}$, as needed.

For our fixed profile $\profile$, we now have that each $\Cov[\propdistrankingi]$ has minimum eigenvalue at least $\frac{\minprob(S_\phi)}{m!}$. Since the minimum eigenvalue of the sum of matrices is at least the sum of the minimum eigenvalues of each matrix, the minimum eigenvalue of $\sum_{i =1}^n\Cov[\propdistrankingi]$ is at least $n \cdot \frac{\minprob(S_\phi)}{m!}$. Finally, to get $\Cov[\propdistprof]$, we scale this sum down by $n^2$ which scales the minimum eigenvalues equivalently, yielding a minimum eigenvalue of at least $\frac{\minprob(S_\phi)}{nm!}$. Note that the minimum eigenvalue here is positive, meaning $\Cov[\propdistprof]$ has no zero eigenvalues, meaning it is invertible. \qed
\end{proof}

\noindent \textbf{Proof of \Cref{lem:normality}.}
Fix $\noisemodel, \phi,$ and $\profile \in \profiles_n$.
Since $\propdistprof = \nicefrac{1}{n} \sum_{i = 1}^n \propdistrankingi$ where each of these summands is independent, our goal will be to apply a version of the Berry-Esseen bound, as stated below:

\begin{lemma}[Restatement of Berry-Esseen as in~\cite{bentkus2005lyapunov}]\label{lem:berryesseen} Let $Y_1,\dots,Y_n$ be independent, mean-zero, $\R^{m!-1}$-valued random variables. Let $S = Y_1+\cdots+Y_n$, and let $C^2$ be the covariance matrix of $S$, assumed invertible. Let $\mathcal{N}(0,C^2)$ be a $m!-1$-dimensional Gaussian with mean zero and covariance $C^2$. Then for any convex subset $X \subseteq \R^{m!-1}$,
\[|\Pr[S \in X] - \Pr[\mathcal{N}(0,C^2) \in X]| \leq O((m!-1)^{1/4}) \cdot \left(\sum_{i = 1}^n \E[|C^{-1}Y_i|^3]\right).\]
\end{lemma}

In order to apply the Berry-Esseen bound, we use the properties of the covariance matrix $\Cov[\propdistprof]$ from \Cref{lem:covariance}.

By \Cref{lem:covariance}, $\Cov[\propdistprof]$ is invertible, so we can apply the Berry-Esseen bound. For consistency with the form of the stated bound, we first translate both our distribution, $\propdistprof$, and the Gaussian to which we want to show it converges, to make both mean-zero. That is, we will show the equivalent statement about $\propdistprof - \E[\propdistprof]$ approaching $\mathcal{N}(0, \Cov[\propdistprof])$. Note that subtracting the expectations of both $\propdistprof$ and $\mathcal{N}(\E[\propdistprof], \Cov[\propdistprof])$ translates the the distributions identically, and note that the convexity of the quantified sets $X$ is invariant under translation. Therefore, proving the claim on the translated version of our distribution implies the claim on our original distribution.

By linearity of expectation, we can express our translated distribution as the sum of $n$ independent random variables:
$$\propdistprof - \E[\propdistprof] = \sum_{i = 1}^n \nicefrac{1}{n} (\propdistrankingi - \E[\propdistrankingi]).
$$
We let $Y_i$ be the random variable distributed as a single term of the above sum, $\nicefrac{1}{n} (\propdistrankingi - \E[\propdistrankingi])$. Note that $Y_i$ has mean zero, and covariance $\nicefrac{1}{n^2} \cdot  \Cov[\propdistrankingi]$. 

Now, all that remains to show is that the following bound on the convergence rate holds:
$$O((m! - 1)^{1/4}) \cdot \left(\sum_{i = 1}^n \E\left[\left|\Cov[\propdistprof]^{-1/2} Y_i\right|^3\right] \right) \le  \frac{O((m! - 1)^{1/4})}{(\lambda^{min, \noisemodel, \phi})^{3/2}} \cdot \frac{1}{\sqrt{n}}.$$
We note that the exponentiated $\Cov[\propdistprof]^{-1/2}$ in this expression is well defined because the matrix is symmetric.

We first show that $|Y_i| \leq 2/n$ for all $i$, where $|Y_i|$ denotes the $L_2$ norm of $Y_i$. Recall that $\propdistrankingi$ is always either a basis vector or the all $0$s vector, and $\E[\propdistrankingi]$ is the vector whose entry corresponding to $\ranking'$ is $\Pr[\rankingdistranking = \ranking']$. 
Then, after subtracting the second vector from the first, the negative entries in the resulting vector can sum in magnitude to at most the sum of these probabilities, which is at most $1$. Similarly, the positive entries can also sum to at most $1$. hence, the $L_1$ norm before scaling by $1/n$ is at most $2$. Using the fact that $L_2$ norms are at most $L_1$ norms, we get that this continues to hold for the $L_2$ norm. After dividing by $n$, we get that $|Y_i| \le 2/n$, as needed.

Now, per \Cref{lem:covariance}, $\Cov[\propdistprof]$ has minimum eigenvalue at least $\frac{\minprob(S)}{m!n}$. It therefore holds that $\Cov[\propdistprof]^{-1/2}$ has maximum eigenvalue at most $$\left(\frac{\minprob(S)}{m!n}\right)^{-1/2} = \sqrt{\frac{m!n}{\minprob(S)}}.$$ 
Multiplying a vector by a matrix can scale the norm by at most the matrix's maximum eigenvalue. Thus, combined with our observation that $|Y_i| \leq 2/n$, the following bound will always hold:
$$|\Cov[\propdistprof]^{-1/2} Y_i|^3 \le \left(\sqrt{\frac{m!n}{\minprob(S)}} \cdot \frac{2}{n}\right)^3 = \frac{8 \cdot (m!)^{3/2} }{n^{3/2} \cdot \minprob(\noisemodel_\phi)^{3/2}}$$
Because this bound holds deterministically on the term above, it must hold also for the expectation of the term above. Thus, by summing over all $i$, we get that
$$
    \sum_{i = 1}^n \E[|\Cov[\propdistprof]^{-1/2} Y_i|^3] \le \frac{8 \, (m!)^{3/2} }{ \sqrt{n} \cdot \minprob(\noisemodel_\phi)^{3/2}}.
$$
Since $O\left((m! - 1)^{1/4}\right) \cdot 8m!^{3/2} \in O\left((m!)^{7/4}\right)$, multiplying by the Berry-Esseen constant yields the lemma statement.\qed

\subsection{Proof of \Cref{thm:brittle_convex}} \label{app:brittle_convex}
   Fix $\noisemodel$, $C$, and $X = \bigcup_{j = 1}^\ell X_j$ where each $X_j$ is convex, $X$ is measure $0$, and $\prop(\profiles^{\neg C}) \subseteq X$. Since $X$ is measure $0$, this implies each $X_j$ has measure $0$. Fix $\phi \in (0, 1]$, a profile $\profile \in \profiles_n$, and $\phi' \in [\phi, 1]$.
    
    Fix an arbitrary $X_j$. Note that since $X_j$ has measure zero, the probability mass placed on $X_j$ by a Gaussian with invertible covariance matrix is $0$. Hence, \Cref{lem:normality} immediately imply that 
   $\Pr[\prop(\noisemodel_{\phi'}(\profile)) \in X_j] - 0 \le \frac{ O((m!)^{7/4})}{\sqrt{n} \cdot \minprob(\noisemodel_{\phi'})^{3/2}}$. Using the monotonicity of $\minprob$ (Assumption~\ref{ass:monotonicity}), we get that this is at most $\frac{ O((m!)^{7/4})}{\sqrt{n} \cdot \minprob(\noisemodel_{\phi})^{3/2}}$ (with $\phi'$ replaced with $\phi$). Union bounding over all $\ell$ sets $X_j$ tells us that
   $$\Pr[\prop(\noisemodel_{\phi'}(\profile)) \in X] \le \frac{\ell \cdot O((m!)^{7/4})}{\sqrt{n} \cdot \minprob(\noisemodel_{\phi})^{3/2}},$$
   as needed. \qed

\subsection{Proof of Theorem~\ref{thm:generalized_brittleness}} \label{app:gen_brittleness}
\begin{proof}
    Fix $\noisemodel,$ $\phi$, and $\delta(n)$. Next, fix a number $n$, a hyperplane $G \in \mathcal{G}$, a $\phi' \in [\phi, 1]$, and a $\profile^\star \in \profiles_n$. Here, $G$ is a hyperplane in $m!-1$ dimensions, meaning it is defined by a linear equation whose variables are the profile proportions for rankings in $\rankingsmissing$, each weighted by a coefficient, $a_\ranking$, along with a constant, $b$. In other words,
    it is the set of proportions $\proportions$ satisfying
    \begin{equation} \label{eqn:hyperplane}
        b = \sum_{\ranking \in \rankingsmissing} a_\ranking \cdot \proportion_\ranking. 
    \end{equation}
    Formally, we need to consider only the intersection of $G$ with our convex hull $H$, but for our results, it is irrelevant whether we formally make this restriction.
    
    We will upper bound $\Pr[d(\prop(\noisemodel_{\phi'}(\profile^\star)), G) \le \delta(n)]$ by an $o(1)$ function of $n$ (allowing this convergence rate to depend on $\phi$, $m$, and $\noisemodel$, but not $G$, $\phi'$, or $\profile^\star$). 
    
    Recall that the hyperplane $G$ has a coefficient for each ranking except the ``missing'' ranking, $\ranking_{-1}$. We choose $\ranking^{max} \in \rankings$ such that $\ranking^{max}$ is the ranking corresponding to a largest-magnitude coefficient in the definition of $G$, i.e., $\ranking^{max} \in \argmax_{\ranking \in \rankingsmissing} |a_\ranking|$. Now, we will define two types of events.

    First, let $\event^{few} = \set{\profile \in \profiles_n \suchthat |\set{i \in N \suchthat \ranking_i \in \set{\ranking^{max}, \ranking_{-1}}}| < np}$ --- that is, $\event^{few}$ is the set of profiles in which fewer than $np$ agents end up voting either $\ranking^{max}$ or $\ranking_{-1}$ in $\prop(\noisemodel_{\phi'}(\profile^\star))$.
    
    Let $\mathcal{V}^{\ge np} = \set{V \subseteq N \suchthat |V| \ge np}$ be the collection of all sets of agents of size at least $np$. For a set of such agents $V \in \mathcal{V}^{\ge np}$, we denote the complement of this set of agents as $\overline{V} = N \setminus V$. Now, slightly abusing notation, define $\widetilde{\profiles}^{\overline{V}} = \set{(\ranking_i)_{i \in \overline{V}} \suchthat \ranking_i \in \rankings \setminus \set{\ranking^{max}, \ranking_{-1}}}$ to be the set of all \emph{partial} profiles in which all agents in $\overline{V}$ have a ranking other than $\ranking^{max}$ or $\ranking_{-1}$. For $V \in  \mathcal{V}^{\ge np}$ and partial profile $\widetilde{\profile} \in \widetilde{\profiles}^{\overline{V}}$, let $\event^{V, \widetilde{\profile}}$ be the event that agents in $V$ either vote $\ranking$ or $\ranking'$, and all other agents vote as in $\widetilde{\profile}$, that is, 
    $$\event^{V, \widetilde{\profile}} = \set{\profile \in \profiles \suchthat \ranking_i = \widetilde{\ranking}_i \text{ for } i \in \overline{V} \text{ and } \ranking_i \in \set{\ranking^{max}, \ranking_{-1}} \text{ for } i \in V}.$$ 
    
    In the remainder of the proof, we will use that the event $\event^{few}$, along with the set of all events of the form $\event^{V,\widetilde{\profile}}$ (that is, for all subsets of agents $V \in \mathcal{V}^{\ge np}$ and partial profiles $\widetilde{\profile} \in \widetilde{\profiles}^{\overline{V}}$), form a partition of the full space of profiles. To see how we will use this, recall that we are trying to show that $d(\prop(\noisemodel_{\phi'}(\profile^\star),G)$ is likely to be somewhat large. We will show this by showing that, conditioned on any individual event $\event^{V,\widetilde{\profile}}$, the probability of this distance being large is fairly high, and the remaining event, $\event^{few}$ is unlikely to occur.
    
    We will show two claims:
    \begin{enumerate}
        \item There is some $f(n) \in o(1)$ such that
    $$\Pr\left[\event^{few}\right] \le f(n).$$
    \item There is some $g(n) \in o(1)$ such that for all $\event^{V, \profile}$,
    $$\Pr\left[d(\prop(\noisemodel_{\phi'}(\profile^\star)), G) \le \delta(n) \conditioned \event^{V, \widetilde{\profile}}\right] \le g(n).$$
    \end{enumerate}
    We first show that together these are sufficient to prove the bound. Indeed, by the law of total probability, as these events form a partition
    \begin{align*}
        \Pr\left[[d(\prop(\noisemodel_{\phi'}(\profile^\star)), G) \le \delta(n)\right]
        &= \Pr\left[d(\prop(\noisemodel_{\phi'}(\profile^\star)), G) \le \delta(n) \conditioned \event^{few}\right]\cdot \Pr\left[\event^{few}\right]\\
        &\qquad + \sum_{\event^{V, \profile}} \Pr\left[d(\prop(\noisemodel_{\phi'}(\profile^\star)), G) \le \delta(n) \conditioned \event^{V, \profile}\right]\cdot \Pr\left[\event^{V, \profile}\right]\\[0.75em]
        &\le \Pr\left[d(\prop(\noisemodel_{\phi'}(\profile^\star)), G) \le \delta(n) \conditioned \event^{few}\right]\cdot f(n) +\sum_{\event^{V, \profile}} g(n) \cdot \Pr\left[\event^{V, \profile}\right] \\
        &=\Pr\left[d(\prop(\noisemodel_{\phi'}(\profile^\star)), G) \le \delta(n) \conditioned \event^{few}\right]\cdot f(n) +  g(n) \sum_{\event^{V, \profile}} \Pr\left[\event^{V, \profile}\right]\\
        &\le  1 \cdot f(n) + g(n) \cdot 1\\
        &= f(n) + g(n) \in o(1).
    \end{align*}
    
    We now show the claims. The first claim follows from a straightforward Chernoff bound: each of the $n$ agents places probability at least $2p$ on ending up in either $\ranking^{max}$ or $\ranking_{-1}$, hence the expected number of agents with either $\ranking^{max}$ or $\ranking_{-1}$ is at least $2np$. Then, given that agents' rankings are sampled independently, the probability of the total number of such agents being less than half of this expectation is exponentially small in $n$.
    
    We now show the second claim. 
    Fix an arbitrary $\event^{V, \widetilde{\profile}}$ and let us consider the conditional distribution of $\prop(\noisemodel_\phi(\profile^\star))$ conditioned on this event. Note that all proportions in the support of this distribution match on all entries except that of $\ranking^{max}$. Let $\widetilde{\proportions}_{-\ranking^{max}}$ be the partial proportions over all these matching entries, i.e., all rankings except $\ranking^{max}$, 
   so that $\proportions_{-\ranking^{max}} = \widetilde{\proportions}_{-\ranking^{max}}$ for all $\proportions \in \prop(\event^{V, \widetilde{\profile}})$. 
    
    Now, since $a_{\ranking^{max}} \ne 0$, by \Cref{eqn:hyperplane}, there exists exactly one completion of the partial proportions $\proportions_{-\ranking^{max}}$ such that the resulting proportion lies on $G$. Namely, 
    $$
        p^{V, \widetilde{\profile}} := \frac{b - \sum_{\ranking \in \rankings \setminus \set{\ranking^{max}, \ranking_{-1}}} a_\ranking \cdot \tilde{\proportion}_\ranking}{a_{\ranking^{max}}}.
    $$
    
    Note that this completion need not be a valid proportion, as the value assigned to the proportion $\ranking^{max}$, $p^{V, \widetilde{\profile}}$, could need to be irrational, negative, or make all proportions add up to strictly more than one.

    Now, we consider all completions of $\proportions_{-\ranking^{max}}$ which assign values to the proportion of ranking $\ranking^{max}$ at least $\delta(n)$ larger or smaller than $p^{V, \widetilde{\profile}}$. In particular, we will show (1) that all such proportions are at least $\delta(n)$ $L_1$ distance from all points on $G$, and (2) that we are likely to draw such a profile from $\prop(\noisemodel_\phi(\profile^\star))$ conditioned on $\event^{V, \widetilde{\profile}}$. 
    
    Fix such a completion of $\widetilde{\proportions}_{-\ranking^{max}}$, call it $\proportions$, so that $|\proportion_{\ranking^{max}} - p^{V, \widetilde{\profile}}| > \delta(n)$. By this assumption, we have the following, where in the first step we replace $b$ according to \Cref{eqn:hyperplane}:
    \begin{align*}
    \left|\sum_{\ranking \in \rankingsmissing} a_\ranking \proportion_\ranking - b \right|
    &= \left|\sum_{\ranking \in \rankingsmissing} a_\ranking \proportion_\ranking - \left(p^{V, \widetilde{\profile}} \cdot a_{\ranking^{max}} + \sum_{\ranking \in \rankings \setminus \set{\ranking^{max}, \ranking_{-1}}} a_\ranking \proportion_\ranking \right)\right| \\
    &= |a_{\ranking^{max}} \cdot \proportion_{\ranking^{max}}  - p^{V, \widetilde{\profile}} \cdot a_{\ranking^{max}}|\\
    &> |\delta(n) \cdot a_{\ranking^{max}}|
    \end{align*}
    Since $a_{\ranking^{max}}$ is maximal in magnitude over all coefficients in $G$, the $\delta(n)$-radius $L_1$-ball around $\proportions$ does not intersect $G$, because changing any entry of $\proportions$ by at most $\delta(n)$ cannot change $\sum_{\ranking \in \rankingsmissing} a_\ranking p_\ranking$ by more than $\delta(n) \cdot a_{\ranking^{max}}$.
    
    We have shown that all completions of $\proportions_{-\ranking^{max}}$ assigning values of $\ranking^{max}$ outside $p^{V, \widetilde{\profile}} \pm \delta(n)$ must be more than $\delta(n)$ $L_1$ distance from $G$. Next, we show that the probability of drawing such a completion is likely. We will do this by upper-bounding the probability placed on $\proportion_{\ranking^{max}}$ falling in the interval $p^{V, \widetilde{\profile}} \pm \delta(n)$ by showing its distribution is approximately normal using the Berry-Esseen bound, and showing the normal distribution does not place much mass on this small interval. 
    
    To this end, let $I_i$ be the indicator random variable that agent $i \in V$ chooses $\ranking^{max}$ (if they do not choose $\ranking^{max}$, they choose $\ranking_{-1}$). 
    Let $S = \sum_{i \in V} I_i$ be the random variable representing the total number of agents in $V$ that vote for $\ranking^{max}$. Note that the resulting proportion $\prop_{\ranking^{max}}$ will be in the range $p^{\overline{V}, \widetilde{\profile}} \pm \delta(n)$ iff $S$ is in the range $n(p^{\overline{V}, \widetilde{\profile}} \pm \delta(n))$, an interval of size $2n \delta(n)$. We now show that the probability $S$ is in \emph{any} interval of size $2n \delta(n)$ is $o(1)$. 
    
    To do this, we first state the Berry-Esseen bound.
    \begin{lemma}[Berry-Esseen Bound~\cite{berry1941accuracy,esseen1942liapunov}]\label{lem:berry1d}
    Let $X_1, \ldots, X_n$ be independent random variables with $\E[X_i] = \mu_i$, $\E[|X_i^2 - \mu_i|] = \sigma_i^2 > 0$, and $\E[|X_i^3 - \mu_i|] = \rho_i < \infty$. Let $T = X_1 + \cdots + X_n$. Let $F$ be the cdf of $\frac{T - \sum_{i = 1}^n \mu_i}{{\sqrt{\sum_{i = 1}^n\sigma_i^2}}}$. Then, there exists an absolute constant $C_1$ such that $$\sup_{x \in \R} |F(x) - \Phi(x)| \le C_1 \cdot \left(\sum_{i = 1}^n \sigma_i^2\right)^{-1/2} \cdot \max_{1 \le i \le n} \frac{\rho_i}{\sigma_i^2}.$$
\end{lemma}
We would like to directly apply \Cref{lem:berry1d} to our $S$, where each $X_i = I_i$. We first consider the quantity $\sigma_i^2$ for each $i$.
    Note that $\minprob(\noisemodel_\phi') \le \Pr[I_i = 1] = 1-\Pr[I_i = 0] \le 1 - \minprob(\noisemodel_\phi')$. Further, by monotonicity (\Cref{ass:monotonicity}), $\minprob(\noisemodel_\phi') \ge \minprob(\noisemodel_\phi')$, so $\minprob(\noisemodel_\phi) \le \Pr[I_i = 1] \le 1 - \minprob(\noisemodel_\phi)$. Since each $I_i$ is a Bernoulli, each $\sigma_i^2 \ge \minprob(\noisemodel_\phi) \cdot (1 - \minprob(\noisemodel_\phi)) \ge \nicefrac{\minprob(\noisemodel_\phi)}{2}$. This implies that $\sum_{i = 1}^n \sigma_i^2 \ge n \cdot \nicefrac{\minprob(\noisemodel_\phi)}{2}$. Further, note that since each $I_i$ is bounded in $[0, 1]$, each $\rho_i \le 1$. This implies the error bound of \Cref{lem:berry1d} is at most
    $$
    C_1 \cdot \left(n \cdot \frac{\minprob(\noisemodel_\phi)}{2}\right)^{-1/2} \cdot \frac{2}{\minprob(\noisemodel_\phi)} = O\left(\frac{1}{\minprob(\noisemodel_\phi)^{3/2} \cdot \sqrt{n}} \right).
    $$
    Further, the probability $S$ is in some range of size $2n \delta(n)$ will be equivalent to the probability $\frac{S - \sum_{i = 1}^n \mu_i}{\sqrt{\sum_{i = 1}^n \sigma_i^2}}$ is in a specific range of size $\frac{2 n\delta(n)}{\sqrt{\sum_{i = 1}^n \sigma_i^2}} \le \frac{2\delta(n)\sqrt{2n}}{\sqrt{\minprob(S_\phi)}}$. Let $F$ be the CDF of this random variable. The probability this random variable is in a range of this size is at most
    \begin{align*}
        &\sup_x \left(F\left(x + \frac{2\delta(n)\sqrt{2n}}{\sqrt{\minprob(S_\phi)}}\right) - F(x) \right)\\
        \le & \sup_x \left(\Phi\left(x + \frac{2\delta(n)\sqrt{2n}}{\sqrt{\minprob(S_\phi)}}\right) - \Phi(x) \right) + 2 O\left(\frac{1}{\minprob(\noisemodel_\phi)^{3/2} \cdot \sqrt{n}} \right)\\
       = & \sup_x  \left( \int_x^{x + \frac{2\delta(n)\sqrt{2n}}{\sqrt{\minprob(S_\phi)}}} \phi(x) \right)  + 2O\left(\frac{1}{\minprob(\noisemodel_\phi)^{3/2} \cdot \sqrt{n}} \right) \\ 
       \le & \sup_x  \left( \int_x^{x + \frac{2\delta(n)\sqrt{2n}}{\sqrt{\minprob(S_\phi)}}} \frac{e}{\sqrt{2 \ranking}} \right)  + 2O\left(\frac{1}{\minprob(\noisemodel_\phi)^{3/2} \cdot \sqrt{n}} \right)\\ 
       = & \sup_x  \left( \frac{e}{\sqrt{2 \ranking}} \cdot \frac{2\delta(n)\sqrt{2n}}{\sqrt{\minprob(S_\phi)}} \right)  + 2O\left(\frac{1}{\minprob(\noisemodel_\phi)^{3/2} \cdot \sqrt{n}} \right)\\
       = & O\left(\frac{1}{\sqrt{\minprob(S_\phi})} \cdot \delta(n)\sqrt{n} \right) + O\left(\frac{1}{\minprob(\noisemodel_\phi)^{3/2} \cdot \sqrt{n}}\right)\\
       \in & o(1).
    \end{align*}
    This is our desired $g(n)$ which completes the proof of the second claim.
\end{proof}

\newpage
\section{Supplemental Materials for \Cref{sec:beyond-minprob}}

\subsection{Proof of \Cref{lem:mallows-ubs}}\label{app:mallows-ubs}
\begin{proof}
We will prove these bounds by lower-bounding $q(k)$.
    First, we have that
\[
    q(k) = \frac{1}{1-\phi^{k+1}}\left(1 - \frac{(1-\phi)k\phi^k}{1 - \phi^k}\right) \ge 1 - \frac{(1-\phi)k\phi^k}{1 - \phi^k} \ge 1 - k\phi^k.
\]  
Second, by the AM-GM inequality, we have that
\begin{align*}
    q(k) = \frac{1}{1-\phi^{k+1}}\left(1 - \frac{(1-\phi)k\phi^k}{1 - \phi^k}\right) \ge 1 - \frac{(1-\phi)k\phi^k}{1 - \phi^k}.
\end{align*}
Simplifying, this is equal to $1 - \frac{k\phi^k}{\sum_{j = 0}^{k - 1}\phi^j}$, and the second upper-bound follows:
\begin{align*}
    1 - \frac{k\phi^k}{\sum_{j = 0}^{k - 1}\phi^j} = 1 - \frac{1}{\frac{1}{k}\sum_{j = 1}^{k} \frac{1}{\phi^j}} \ge 1 - \frac{1}{\left(\prod_{j = 1}^{k} \frac{1}{\phi^j}\right)^{1/k}} = 1 - \phi^{\frac{k(k + 1)}{2} \cdot \frac{1}{k}} = 1 - \phi^{\frac{k + 1}{2}} \ge 1 - \phi^{k/2}.
\end{align*}
Third,
\[
    q(k) \ge q(1) = \frac{1}{1 - \phi^2}\left(1 - \frac{(1 - \phi)\phi}{1 - \phi}\right) = \frac{1 - \phi}{1 - \phi^2} = \frac{1}{1 + \phi} = 1 - \frac{\phi}{1 + \phi}.\qedhere
\]
\end{proof}

\subsection{Proof of \Cref{lem:borda-mallows}} \label{app:borda-mallows}
\begin{proof}
We begin with \textsc{Resolvability} and show $\rho(n)-$\textsc{Group-Stability} later.
    Fix a starting profile $\profile$ and two candidates, $a$ and $b$. Our approach will be to first upper bound the probability that $a$ and $b$ have the same Borda score post-noise, and then union bound over all pairs of candidates. For a candidate $c$, let $X^c_i \in \{0,\dots,m-1\}$ be the random variable representing the number of Borda points given to candidate $c$ by voter $i$, post-noise. Let $Y_i = X^a_i - X^b_i$ be the difference in Borda points given to $a$ and $b$ by voter $i$. Our goal is to upper bound $\Pr[\sum_i Y_i = 0]$. As in the proof for \textsc{Plurality}, we will do so via use Berry-Esseen using that the distribution of $\sum_i Y_i$ converges to a Gaussian. Since the Gaussian places zero mass on the value zero, we simply upper bound $\Pr[\sum_i Y_i = 0]$ by the convergence rate of its distribution to a Gaussian, given by Berry-Essen exactly as in \Cref{eq:berry-esseen-ub}.
    
    Again, bounding this upper bound's component terms: 
    Note that $Y_i \in [-(m - 1), (m - 1)]$, so $|Y_i - \mathbb{E}[Y_i]| \le 2m$, implying the same bound on $\max_i \frac{\rho_i}{\sigma_i^2}$. To lower bound each $\sigma_i^2$, we consider the distribution of $Y_i$. In particular, we will upper bound the probability of any specific (integer) value by $\frac{1}{1 + \phi}$. That is, for all values $\ell$,
    \begin{equation}\label{eq:mode-bound}
        \Pr[Y_i = \ell] \le \frac{1}{1 + \phi}
    \end{equation}
    Once we have shown Inequality~\eqref{eq:mode-bound} holds, note that at most one value $\ell$ can be within $1/2$ of $\E[Y_i]$, so this will imply that with probability at least $1 - \frac{1}{1 + \phi} = \frac{\phi}{1 + \phi} \ge \frac{\phi}{2}$, $|Y_i - \mathbb{E}[Y_i]| \ge 1/2$. This allows us to lower bound the variance by $ \frac{\phi}{2} \cdot (\frac{1}{2})^2 = \frac{\phi}{8}$. This gives us an overall bound of $O(\frac{m}{\sqrt{n\phi}})$, and therefore $O(\frac{m^3}{\sqrt{n\phi}})$ after a union bound.

    Finally, we show Inequality~\eqref{eq:mode-bound}. Fix a value $\ell$. Let $\tau \in \mathcal{L}(M \setminus \set{a})$ be an arbitrary partial ranking of the candidates without $\ell$. We will write $\sigma \setminus \set{a} = \tau$ when the partial ranking of $\sigma$ without $a$ included matches $\tau$. We will show that when $\sigma$ is sampled from a Mallows model, conditioned on it matching $\tau$, the probability $a$ is $\ell$ positions above $b$ is at most $\frac{1}{1 + \phi}$. Since $\tau$ was arbitrary, by law of total probability, this implies the same bound in general. Indeed, note that once $\tau$ is fixed, there are $m$ possible rankings after the insertion of $a$. At most one can correspond to a difference of $\ell$. Call this ranking $\sigma$. Further, consider inserting $a$ in the neighboring position. We will call this ranking $\sigma'$. Note that the Kendall tau distance from the starting ranking $\sigma_i$ of $\sigma$ and $\sigma'$ can differ by at most $1$. This implies that $\Pr[\sigma' | \sigma \setminus \set{a} = \tau] / \Pr[\sigma | \sigma \setminus \set{a} = \tau] \ge \phi$. This directly implies that $\Pr[\sigma | \sigma \setminus \set{a} = \tau] \le \frac{1}{1 + \phi}$, as needed.

    For $\rho(n)-$\textsc{Group-Stability}, note that a single voter can affect the difference in scores of a candidate by at most $2(m - 1)$. Hence, as long as the maximal candidate is winning by at least $2(m - 1)\rho(n)$, no group of size $\rho(n)$ can change the outcome. Rather than just doing the analysis for $\sum_i Y_i = 0$, we can do the same for all values in $[-2(m - 1)\rho(n), 2(m - 1)\rho(n)]$, which, via a union bound, adds an additional $O(m\rho(n))$ factor.
\end{proof}

\subsection{Proof of \Cref{lem:minimax-mallows}} \label{app:minimax-mallows}

\begin{proof}
We begin with \textsc{Resolvability}.    
Fix some $n$ that is divisible by $4$ and first, assume $m$ is even (we will describe how to extend it to the odd case later). Consider the following instance. Let $a$ and $b$ be two candidates and let $C$ and $D$ be a partition of the remaining candidates into two sets of equal size, each having $m/2 - 1$ candidates. Let $\tau^C$ and $\tau^D$ be arbitrary orders of the candidates in $C$ and $D$ respectively, and let $\text{rev}(\tau^C)$ and $\text{rev}(\tau^D)$ be the reversals of these rankings respectively. Finally, assume the voter rankings are as follows, each occurring $n/4$ times: 
\begin{enumerate}
    \item $b \succ \text{rev}(\tau^C) \succ a \succ \tau^D$
    \item $b \succ \text{rev}(\tau^D) \succ a \succ \tau^C$
    \item $a \succ \text{rev}(\tau^C) \succ b \succ \tau^D$
    \item $a \succ \text{rev}(\tau^D) \succ b \succ \tau^C$
\end{enumerate}
We will show that except for with exponentially small probability, $a$ and $b$ still beat every other candidate in a pairwise competition. However, with a reasonably large probability, $a$ and $b$ are pairwise tied. When these events occur, $a$ and $b$ have the same optimal Minimax score of $n/2$, while all others have strictly less, and are hence tied.

For the first, consider an arbitrary candidate $c \in C$. The arguments for if $c \in D$ or with $b$ instead of $a$ will be symmetric. Let $X_i$ be a random variable that is $1$ if $i$ ranks $a$ above $c$ in the final ranking and $-1$ otherwise. We would like to upper bound the probability that $\sum_i X_i \le 0$. We will do this using an application of Hoeffding's inequality. To do this, we must obtain bounds on $\mathbb{E}[\sum_i X_i]$.

Suppose $c$ is ranked $k$'th in $\tau^C$. Note that $c$ is only preferred to $a$ on voters in the first group. For such a voter $i$, as there is a probability $\bar{q}(k)$ of a swap, so $\E[X_i] = 2\bar{q}(k) - 1$. For voters in the second group, $a$ is preferred to $c$ by the same magnitude, hence, for such a voter $i$, $\E[X_i] = 1 - 2\bar{q}(k)$, the exact negative of the first.  For voters in the third group, since $a$ is preferred to $c$, they retain their order with probability at least $1/2$, so for such voters $i$, $\E[X_i] \ge 0$. For voters in the fourth group, $a$ is preferred to $c$ by $|D| + 1 + k$ positions, hence, $\E[X_i] \ge 2 q(|D| + 1 + k) - 1$. Since $\bar{q}$ is decreasing and $|D| + 1 + k \ge m/2$, this is at least $1 - 2\bar{q}(m/2)$. Putting this together, we have that \[\E[\sum_i X_i] \ge n/4*(2q(k)  - 1) + n/4(1 - 2q(k)) + n/4(1 - 2\bar{q}(m/2)) = n/4(1 - 2\bar{q}(m/2)).\]
Plugging this into Hoeffding's inequality, since the variables take on values in $[-1, 1]$, this implies that
\[
    \Pr[\sum_i X_i \le 0] \le \exp\left(-\frac{n(1 - 2\bar{q}(m/2))^2}{32} \right).
\]

Union bounding over all $m$ candidates choices of $c$ and $2$ choices of $a$ and $b$, we get that the probability $a$ and $b$ pairwise beat all other candidates except for with probability $2m\exp\left(-\frac{n(1 - 2\bar{q}(m/2))^2}{32} \right)$.

Next, we consider the relative rankings of $a$ and $b$. Note that on groups one and two, $b$ is beating $a$, and on groups three and four, $a$ is beating $b$. In all cases, the margin of victory is $m / 2$ positions. The probability of a swap on any of these rankings is $\bar{q}(m/2)$. Hence, the number of swaps on the first $n/2$ candidates follows a $\text{Bin}(n/2, \bar{q}(m/2))$ distribution, and symmetrically, so does the number of swaps on the second $n/2$. Candidates $a$ and $b$ will be tied in a pairwise competition if the two (independent) draws from these binomials are equal. We show next that this occurs with probability $\Omega(\frac{1}{\sqrt{n\bar{q}(m/2)}})$ which follows directly from the following lemma.

\begin{lemma}
    The probability two independent draws from a $\text{Bin}(n, p)$ distribution are equal is $\Omega(\frac{1}{\sqrt{np}})$. 
\end{lemma}
\begin{proof}
    Fix an arbitrary $n$ and $p$. Note that the mean of a binomial distribution is $np$. By standard Chernoff bounds, if $X$ follows a $\text{Bin}(n, p)$ distribution, $\Pr[|X - \mu| \ge \delta \mu] \le 2\exp(-\delta^2 \mu / 3)$ for all $\delta \in (0, 1)$ where $\mu=np$ . In particular, plugging in $\delta = 2/\sqrt{\mu}$, we get that with constant probability ($\ge 1 - 2\exp(-4/3) \ge .47$), $X$ is within $2\sqrt{\mu}$ of its mean. Note that there are at most $4\sqrt{np} + 1$ integer values within $2\sqrt{np}$ of the mean. With constant probability ($\ge .47^2$), both binomials will take on one of these values. Conditioned on this event, the probability they are equal is at least $\frac{1}{4\sqrt{np} + 1}$. Indeed, suppose conditioned on being one of these $\frac{1}{4\sqrt{np} + 1}$ values, each was taken with probability $p_1, p_2, \ldots, p_{4\sqrt{np} + 1}$ with $\sum_{i = 1}^{4\sqrt{np} + 1} p_i = 1$. The probability they are equal is $\sum_{i = 1}^{4\sqrt{np} + 1} p^2_i \ge \sum_i \frac{p_i}{4 \sqrt{np} + 1} = \frac{1}{4\sqrt{np} + 1}$. Hence, the probability they are equal is at least $\Omega(\frac{1}{\sqrt{np}})$. 
\end{proof}

Combining the fact that $a$ and $b$ are pairwise tied with probability $\Omega(\frac{1}{\sqrt{n\bar{q}(m/2)}})$ and pairwise beat all others with probability $1 - O\left(\exp(-\frac{n(1 - 2\bar{q}(m/2))^2}{32} \right)$ completes the bound.

To handle $m$ which is odd, we can add a dummy candidate to the end of all the rankings. This candidate wins with even smaller probability, so all of the analysis continues to hold. We simply need to replace $\bar{q}(m/2)$ with $\bar{q}(\floor{m/2})$.

Next, consider $1$-\textsc{Strategy-Proofness}. We will have the following starting profile. Let $a, b, x, y$ be candidates, and let $S_1, S_2, S_3, S_4$ be sets of candidates of size $\floor{\frac{m}{4}}$ (as before, we will put remaining candidates at the bottom, but this will not affect the analysis much).
\begin{center}
\begin{tabular}{ll}
    $(n - 2)/4$ voters: & $a \succ \rev(\tau_{S_1}) \succ x \succ \rev(\tau_{S_2}) \succ b \succ \tau_{S_3} \succ y \succ \tau_{S_4}$\\
    $(n - 2)/4$ voters: & $a \succ \rev(\tau_{S_4}) \succ x \succ \rev(\tau_{S_3}) \succ b \succ \tau_{S_2} \succ y \succ \tau_{S_1}$\\
    $(n - 2)/4$ voters: & $b \succ \rev(\tau_{S_1}) \succ y \succ \rev(\tau_{S_2}) \succ a \succ \tau_{S_3} \succ x \succ \tau_{S_4}$\\
    $(n - 2)/4$ voters: & $b \succ \rev(\tau_{S_4}) \succ y \succ \rev(\tau_{S_3}) \succ a \succ \tau_{S_2} \succ x \succ \tau_{S_1}$\\
    $1$ voter: & $x \succ \rev(\tau_{S_2}) \succ b \succ \rev(\tau_{S_4}) \succ y \succ \tau_{S_3} \succ a \succ \tau_{S_1}$\\
    $1$ voter: & $y \succ \rev(\tau_{S_1}) \succ a \succ \rev(\tau_{S_3}) \succ b \succ \tau_{S_4} \succ x \succ \tau_{S_2}$
\end{tabular}
\end{center}

The analysis will begin similarly to \textsc{Resolvability}, showing that $a$ and $b$ both beat all candidates outside of $\{a, b, x, y\}$ by a pairwise majority with all but exponentially small probability. Then, we will show that it is reasonably likely that the relative orders of $a, b, x,$ and $y$ remain exactly as they are in the starting profile for all voters. Under this condition, we can check that the winner is in fact $b$. Indeed, $b$ at minimum pairwise ties with all others, while for all other candidates, we can find a candidate that strictly pairwise beats them, namely $a$ beats $x$ on $n - 1$ votes, $b$ beats $y$ on $n - 1$ votes, and $y$ beats $a$ on $n/2 + 1$ votes, and the rest are beaten by both $a$ and $b$. However, if the last voter bumps up $a$ to the top of their ranking and pushes $b$ to the bottom to instead report $a \succ y \succ \rev(\tau_{S_1}) \succ \rev(\tau_{S_3}) \succ \tau_{S_4} \succ x \succ \tau_{S_2} \succ b$, $a$ will now pairwise beat all other candidates while $b$ will be beaten by $x$ on $n/2 + 1$ votes. This leads $a$ to be the unique winner instead of $b$, an improvement according to manipulating voter.

We now compute the relevant probability bounds. One can check as we did with resolvability that for any candidate $j \in S_i$, even if they beat one of $a$ or $b$ on a certain number of votes, these can be matched with an equal number where $a$ and $b$ beat them by the same amount. For example, consider a candidate $j \in S_1$ ranked $k$'th in $\tau_{S_1}$. Although they beat $a$ on the third group of voters, this can be matched to the voters in the fourth group that prefer $a$ to $j$ by the exact same margin. Further, although the last voter prefers $j$ to $a$, the second to last prefers $a$ to $j$ by the same margin. As before, there is an additional $(n-2)/4$ size group that prefers $a$ to $j$ by a margin of at least $\floor{m/2}$. Hence, by the same analysis as for resolvability, we get that $a$ and $b$ strictly pairwise beat all other candidates except for with probability $2m\exp\left(\frac{-(n - 2)(1 - 2\bar{q}(\floor{m/2}))^2}{32} \right)$.

Next, we consider the probability that an individual voter maintains their order of $\set{a, x, b, y}$. Suppose, without loss of generality, they rank $a \succ x \succ b \succ y$. Note that as long as they continue to rank $a \succ x$, $x \succ b$, and $b \succ y$, then by transitivity, the entire partial ranking will remain. Each of these occurs with probability $1 - q(\floor{m/4})$ as there is always a gap at least this large between the candidates. Hence, union bounding over all voters, we get that all will maintain their orders with probability at least $1 - 3nq(\floor{m/4})$. Putting this together we have that the overall convergence rate is lower bounded by $1 - 3nq(\floor{m/4}) - O\left(m\exp\left(\frac{-(n - 2)(1 - 2\bar{q}(\floor{m/2}))^2}{32} \right)\right)$.
\end{proof}

\section{Supplemental Material from \Cref{sec:discussion}}
\begin{corollary}[Gibbard-Satterthwaite] \label{cor:GS}
Per \Cref{tab:classification2}, there exist several voting rules that are simultaneously non-dictatorial, permit $m\geq 3$, and smoothed-satisfy \textsc{Strategyproofness}.\footnote{$\sqrt{n}$-\textsc{Group-Strategyproofness}$\implies 1$-\textsc{-Group-Strategyproofness}, i.e., \textsc{Strategyproofness}.}
\end{corollary}

\begin{corollary}[ANR] \label{cor:ANR}
Per \Cref{tab:classification}, there exist several voting rules that simultaneously satisfy \textsc{Anonymity} and \textsc{Neutrality} and smoothed-satisfy \textsc{Resolvability}.
\end{corollary}

\begin{corollary}[Condorcet-Participation] \label{cor:condpart}
Per \Cref{prop:group-stability}, there exist several voting rules that simultaneously satisfy \textsc{Condorcet} and smoothed-satisfy \textsc{Participation}.\footnote{$\sqrt{n}$-\textsc{Group-Participation}$\implies 1$-\textsc{Group-Participation}, i.e., \textsc{Participation}.}
\end{corollary}


\noindent \textit{More general criteria.} Here; we consider the more general ``criterion'' that $\profile$ contains no Condorcet cycle. Note that this is not strictly an axiom, as whether it holds on a profile $\profile$ depends only on the profile itself---not on any voting rule. To be formal, we think of this as a more generic criterion $C : \profiles \to \{\textsf{true},\textsf{false}\}$, representing the true/false statement ``$C$ holds on $\profile$''. We can think of $A(R)$ as an example of such a generic criterion. 
\begin{proposition} \label{prop:condcycles} Let $C$ be the criterion that $\profile$ contains no Condorcet Cycle. $C$ is smoothed-violated.
\end{proposition}
\begin{proof}   
    This proposition is proven using the same procedure as for criterion of the form $A(R)$, as before: we find a profile where all nearby histograms have a Condorcet Cycle, implying the preconditions of \Cref{thm:violation-sufficient}.
  We define this profile to be any consistent with the histogram $\proportions$, in which $1/3$ of voters have $a \succ b \succ c$, $1/3$ of voters have $b \succ c \succ a$, and $1/3$ of voters have $c \succ a \succ b$. This profile has a Condorcet cycle: $a$ pairwise-dominates $b$, $b$ pairwise dominates $c$, and $c$ pairwise dominates $a$.

    Now, we will establish that $\proportions$ satisfies the precondition of \Cref{thm:violation-sufficient}. Fix any histogram $\proportions'$ such that $\|\proportions - \proportions'\|_1 < 1/6$. Assume for the sake of contradiction that $\proportions'$ does not contain a Condorcet cycle. It follows that at least one relative pairwise domination flipped from $\proportions$ to $\proportions'$. All pairwise dominations in $\proportions$ were symmetric, so we wlog consider $a$'s pairwise domination of $b$. In order for $b$ to pairwise dominate $a$, at least $1/2 - 1/3 = 1/6$ of voters must have reversed their relative ranking of $a$ and $b$ from $\proportions$ to $\proportions'$---i.e., $\|\proportions' - \proportions\|_1 \geq 1/6$. 
    This is a contradiction, and we conclude the claim.
\end{proof}

\end{document}